\documentclass[12pt]{article}

\usepackage{graphicx}
\usepackage{fancybox}
\usepackage{latexsym}
\usepackage{epsfig}

\usepackage{amsfonts,amsmath,amssymb}
\usepackage{ascmac}

\newcommand{\om}{\omega}
\newcommand{\al}{\alpha}
\newcommand{\ep}{\epsilon}

\newcommand{\la}{\lambda}
\newcommand{\La}{\Lambda}

\newcommand{\vD}{\varDelta}

\newcommand{\tR}{\widetilde{\mbox{R}}}
\newcommand{\sNS}{\msc{NS}}
\newcommand{\stNS}{\widetilde{\msc{NS}}}
\newcommand{\sR}{\msc{R}}
\newcommand{\stR}{\widetilde{\msc{R}}}

\newcommand{\lb}{\lbrack}
\newcommand{\rb}{\rbrack}

\newcommand{\msc}[1]{\mbox{\scriptsize #1}}
\newcommand{\dsp}{\displaystyle}

\newcommand{\bc}{\mathbb C}
\newcommand{\br}{\mathbb R}
\newcommand{\bz}{\mathbb Z}

\newcommand{\bsz}{\mathbb Z}

\newcommand{\cA}{{\cal A}}

\newcommand{\cO}{{\cal O}}
\newcommand{\cN}{{\cal N}}

\newcommand{\cF}{{\cal F}}

\newcommand{\cS}{{\cal S}}

\newcommand{\cQ}{{\cal Q}}

\newcommand{\cZ}{{\cal Z}}
\newcommand{\cW}{{\cal W}}

\newcommand{\tG}{\widetilde{G}}

\newcommand{\tT}{\widetilde{T}}
\newcommand{\tc}{\tilde{c}}
\newcommand{\tgamma}{\tilde{\gamma}}

\newcommand{\hF}{\widehat{F}}
\newcommand{\hf}{\widehat{f}}
\newcommand{\htf}{\widehat{\tilde{f}}}
\newcommand{\hc}{\hat{c}}
\newcommand{\hH}{\widehat{H}}
\newcommand{\hh}{\widehat{h}}
\newcommand{\hG}{\widehat{G}}

\newcommand{\Th}[2]{\Theta_{#1,#2}}
\renewcommand{\th}{{\theta}}

\newcommand{\ch}[2]{\mbox{ch}^{#1}_{#2}}

\newcommand{\chid}{{\chi_{\msc{dis}}}}

\newcommand{\hchid}{\widehat{\chi}_{\msc{dis}}}

\newcommand{\erf}{\mbox{Erf}}
\newcommand{\erfc}{\mbox{Erfc}}
\newcommand{\sgn}{\mbox{sgn}}

\newcommand{\del}{\partial}

\newcommand{\any}{{}^{\forall}}

\newcommand{\nn}{\nonumber\\}

\def\boxit#1{\vbox{\hrule\hbox{\vrule\kern8pt
\vbox{\hbox{\kern8pt}\hbox{\vbox{#1}}\hbox{\kern8pt}}
\kern8pt\vrule}\hrule}}
\def\mathboxit#1{\vbox{\hrule\hbox{\vrule\kern8pt\vbox{\kern8pt
\hbox{$\displaystyle #1$}\kern8pt}\kern8pt\vrule}\hrule}}

\newcommand {\eqn}[1]{(\ref{#1})}

\makeatletter
\@addtoreset{equation}{section}
\def\theequation{\thesection.\arabic{equation}}
\makeatother

\begin{document}

\begin{titlepage}
 \
 \renewcommand{\thefootnote}{\fnsymbol{footnote}}
 \font\csc=cmcsc10 scaled\magstep1
 {\baselineskip=14pt
 \rightline{
 \vbox{\hbox{\today}
       }}}

 \baselineskip=20pt
 
\begin{center}

{\bf \Large  
Duality in $\cN=4$ Liouville Theory


and 

\vspace{3mm}

Moonshine Phenomena
}

 \vskip 1.2cm
 

\noindent{ \large Tohru Eguchi}\footnote{\sf tohru.eguchi@gmail.com}
 \\

\medskip

{\it Department of Physics and Research Center for Mathematical Physics, \\
Rikkyo University, Tokyo 171-8501, Japan}

\vskip 8mm
\noindent{ \large Yuji Sugawara}\footnote{\sf ysugawa@se.ritsumei.ac.jp}
\\

\medskip

 {\it Department of Physical Sciences, 
 College of Science and Engineering, \\ 
Ritsumeikan University,  
Shiga 525-8577, Japan}
 

\end{center}

\bigskip

\begin{abstract}
We consider the  ${\cal N}=4$ Liouville theory by varying the linear dilaton coupling constant $\cal{Q}$. It is known that at two different values of coupling constant ${\cal Q}=\sqrt{{2\over N}},-(N-1)\sqrt{{2\over N}}$ system exhibits two different small ${\cal N}=4$ superconformal symmetries  with central charge $c=6$ and $c=6(N-1)$, respectively. In the context of string theory  these two theories are considered to describe Coulumb and Higgs branch of the theory and expected to be dual to each other. We study the Mathieu and umbral moonshine phenomena in these two theories and discuss their dual description. We mainly consider the case of $A_N$ type modular invariants.

\end{abstract}


\setcounter{footnote}{0}
\renewcommand{\thefootnote}{\arabic{footnote}}

\end{titlepage}

\baselineskip 18pt

\vskip2cm 


\section{Introduction}

In this paper we study the ${\cal N}=4$ supersymmetric Liouville theory in order to have a deeper understanding on the Mathieu and umbral moonshine phenomena discovered recently   \cite{EOT,umbral1,umbral2}. It is known that when one perturbs the large ${\cal N}=4$ theory by varying the strength of the linear dilaton term ${\cal Q}$ there are two special values of ${\cal Q}$ where the theory possesses the small ${\cal N}=4$ supersymmetry  with the central charges $c=6$ and $c=6(N-1)$. 
These two theories have different $SU(2)_R$ symmetry, i.e. different components of $SO(4)=SU(2)\times SU(2)$ of large ${\cal N}=4$ Liouville theory.  
In the context of string theory these describe Higgs and Coulumb branches of the theory and are expected to have a description dual to each other.

While the values of central charges ($c=6, \, 6(N-1)$) appear to differ greatly at large $N$, 
it is known that the value of the effective central charge does not vary much in Liouville theory as dilaton coupling is varied.   
Thus we have dual pairs of  ${\cal N}=4$ CFT's (case 1 and 2) with comparable degrees of freedom.  
We introduce various assumptions on the elliptic genera of case 1,2 theories and 
derive algebraic identities and inequalities. We can determine the shadows 
and elliptic genera of case 1,2 theories which exhibit nice dual descriptions. 
We also discuss the relation of the work \cite{CH} to 
case 1 theories.

We work mainly with $A$-type modular invariant in this paper; we have some subtleties   
in the case of $D,E$ type modular invariants of umbral moonshine which will be discussed in a 
future publication.


~

\section{Preliminaries : Free Field Realizations of $\cN=4$ Superconformal Systems}
 
In this preliminary section we summarize the basic properties of 
the large/small $\cN=4$ superconformal algebras
and their free field realizations.

~

\subsection{Large $\cN=4$ Superconformal Algebra}

The large $\cN=4$ superconformal algebra (SCA), often 
denoted as `$\cA_{\gamma}$',  is defined as the  
$SO(4) = SU(2)\times SU(2)$ extension of Virasoro 
algebra \cite{STV,Ivanov:1988rt} (see also \cite{GMMS} for a good summary). 
We have a stress-energy tensor $T$, four supercurrents
$G^a$ ($a=0,\ldots,3$), two $SU(2)$ currents $A^{+,i}$,
$A^{-,i}$ ($i=1,\ldots,3$) whose levels are $k^+$, $k^-$, 
one $U(1)$ current $U$, and four Majorana fermions $Q^a$
($a=0,\ldots, 3$).
The unitarity requires that $k^+$, $k^-$ should be positive
integers, and the central charge is given as 
\begin{eqnarray}
 c= \frac{6k^+k^-}{k^++k^-}.
\end{eqnarray}
We set $\gamma :=  \frac{k^-}{k^+ + k^-}$,
which parameterizes the `mixing' of two $SU(2)$ currents. 
The non-trivial part of large $\cN=4$ algebra is written 
as follows\footnote{We are here using a slightly different 
  convention from that of \cite{STV,GMMS}. 
   Our $A^{\pm,i}$, $U$, $Q^a$ correspond to 
   $iA^{\pm,i}$, $iU$, and $iQ^a$ in \cite{STV,GMMS}.};
\begin{eqnarray}
 && G^a(z)G^b(w) \sim \frac{2c}{3}\frac{\delta^{ab}}{(z-w)^3}
 + \frac{8i}{(z-w)^2}\left\{\gamma \al^{+,i}_{ab}A^{+,i}(w)
 + (1-\gamma) \al^{-,i}_{ab}A^{-,i}(w)\right\} \nn
&& \hspace{1cm}
 + \frac{4i}{z-w}\left\{\gamma \al^{+,i}_{ab}\partial
 A^{+,i}(w)
 + (1-\gamma) \al^{-,i}_{ab}\partial A^{-,i}(w)\right\}
 + \frac{2 \delta^{ab}}{z-w} T(w)~, \nn
&& A^{\pm,i}(z)A^{\pm,j}(w) \sim 
\frac{\frac{k^{\pm}}{2}\delta^{ij}}{(z-w)^2} 
+ \frac{i \ep^{ijk}}{z-w} A^{\pm,k}(w) ~, \nn
&& Q^a(z)Q^b(w) \sim \frac{\frac{k^++k^-}{2}}{z-w} \delta^{ab}~,
~~~ U(z)U(w) \sim \frac{\frac{k^++k^-}{2}}{(z-w)^2}~, \nn
&& A^{\pm,i}(z)G^a(0) \sim 
\left(\gamma - \frac{1}{2}\mp \frac{1}{2}\right)
\frac{2 \al^{\pm,i}_{ab}}{(z-w)^2}Q^b(w) + 
\frac{i\al^{\pm,i}_{ab}}{(z-w)}G^b(w)~, \nn
&& A^{\pm,i}(z)Q^a(w) \sim \frac{i\al^{\pm,i}_{ab}}{z-w}Q^b(w)
~, \nn
%
&& Q^a(z)G^b(w) \sim \frac{2}{z-w} \left\{
 \al^{+,i}_{ab}A^{+,i}(w) - \al^{-,i}_{ab}A^{-,i}(w)
\right\} + \frac{\delta^{ab}}{z-w} U(w) ~, \nn
&& U(z)G^b(w) \sim \frac{1}{(z-w)^2} Q^a(w)~,
\label{large N=4}
\end{eqnarray}
where we introduced the $4\times 4$ matrices
\begin{eqnarray}
&& \al^{\pm,i}_{ab} \equiv \frac{1}{2}
\left(\pm\delta_{ia}\delta_{b0} \mp\delta_{ib}\delta_{a0}
+ \ep_{iab}\right)~,
\label{alpha}
\end{eqnarray}
(the third term only contributes if $a,b\neq 0$).
They obey the $SO(4)$ commutation relations;
\begin{eqnarray}
 \lb \al^{\pm,i}, \, \al^{\pm,j} \rb = -\ep^{ijk} \al^{\pm,k}~,
~~~ \lb \al^{\pm,i}, \, \al^{\mp,j} \rb =0~, ~~~
 \{ \al^{\pm,i}, \al^{\pm,j}\} = -\frac{1}{2}\delta^{ij}~.
\end{eqnarray}

~


\subsection{Free Field Realization of Large $\cN=4$ SCA}

Let us consider a conformal system composed of 
a free boson $\phi$, 
four Majorana fermions
$\psi^a$ ($a=0,\ldots, 3$), and $SU(2)_k$ current $j^i$ 
$(i=1,2,3)$. $\phi$ and $\psi^a$ are normalized as 
$\phi(z)\phi(w) \sim -\ln (z-w)$, 
$\psi^a(z)\psi^b(w) \sim \frac{\delta^{ab}}{z-w}$, 
and the $SU(2)_k$ current algebra is written in our convention
as 
\begin{eqnarray}
 j^i(z)j^j(w) \sim \frac{\frac{k}{2}\delta^{ij}}{(z-w)^2}
  + \frac{i \ep^{ijk}}{z-w} j^k(w)~.
\end{eqnarray}
We can combine the fermionic components to the $SU(2)$ currents, and 
obtain the level $N \equiv k+2$ `total currents';
\begin{equation}
J^i(z) := j^i(z) -\frac{i}{2}\ep_{ijk} \psi^j(z) \psi^k(z).
\label{total J}
\end{equation}


We set 
\begin{eqnarray}
&&T := -\frac{1}{2}(\partial \phi)^2 
-\frac{1}{2} \psi^a \partial \psi^a + \frac{1}{N} j^i j^i , \nn
&& G^a := i\psi^a \partial \phi - 2\sqrt{\frac{2}{N}}
 \al^{+,i}_{ab} j^i \psi^b 
- \frac{1}{6} \sqrt{\frac{2}{N}} 
i \ep_{abcd}\psi^b \psi^c \psi^d, 
\nn
&& A^{+,i} := -\frac{i}{2}\al^{+,i}_{ab}\psi^a\psi^b + j^i, 
~~~ A^{-,i} := -\frac{i}{2}\al^{-,i}_{ab} \psi^a\psi^b, 
\nn
&& U:= \sqrt{\frac{N}{2}}i\partial \phi, 
~~~ Q^a := \sqrt{\frac{N}{2}}\psi^a.
\label{ff large N=4}
\end{eqnarray}
More explicitly, we can rewrite 
\begin{eqnarray}
&& G^0 = i\psi^0 \partial \phi + \sqrt{\frac{2}{N}} 
\left(
\psi^i j^i - i \psi^1\psi^2\psi^3\right) ,
\end{eqnarray}
which corresponds to the $\cN=1$ subalgebra
($\psi^0$ is identified as the superpartner of $\phi$), and
also 
\begin{eqnarray}
 && G^i = i\psi^i \partial \phi - \sqrt{\frac{2}{N}}
\left(
\psi^0 j^i+ \ep_{ijk}\psi^j j^k -\frac{i}{2} \ep_{ijk}
\psi^j\psi^k\psi^0
\right), \\
&& A^{+,i} = -\frac{i}{2} \psi^i \psi^0 
-\frac{i}{4}\ep^{ijk}\psi^j\psi^k + j^i~, ~~~
A^{-,i} = \frac{i}{2} \psi^i \psi^0 
-\frac{i}{4}\ep^{ijk}\psi^j\psi^k ~.
\end{eqnarray}
They generate the large ${\cal N}=4$ algebra with parameters
\begin{eqnarray}
 c= \frac{6(N-1)}{N} \left(\equiv 
\frac{6k^+k^-}{k^++k^-}\right), ~~~ k^+=N-1, ~~~ k^-=1, 
~~~ \gamma= \frac{1}{N}.
\label{c large N=4}
\end{eqnarray} 
In fact, the total central charge is calculated as 
\begin{eqnarray}
 c= 1 + 4\times \frac{1}{2} + \frac{3k}{k+2} 
= \frac{6(k+1)}{k+2}\equiv \frac{6(N-1)}{N}.
\end{eqnarray}

As is familiar, the `zero-mode subalgebra' of the large $\cN=4$ is the 
super-Lie algebra $D(2,1;\al)$ with $\al \equiv 
\frac{\gamma}{1-\gamma}
$
generated by $L_0$, $L_{\pm 1}$, $G^a_{\pm 1/2}$,
$A^{\pm,i}_0$, $U_0$, $Q^a_{\pm 1/2}$ (for the NS sector).

~

\subsection{Deformation by Linear Dilaton}

We next consider a deformation of \eqn{ff large N=4}
by turning on the linear dilaton (background charge) along $\phi$.
We shall deform as 
\begin{eqnarray}
 && T(z)~\rightarrow~ \tT(z) :=  T(z) 
 - \frac{\cQ}{2}\partial^2 \phi, \nn
 && G^a(z) ~\rightarrow~ \tG^a(z) :=  G^a(z) 
 + \cQ i \partial \psi^a(z).
\label{deform 1}
\end{eqnarray}
It keeps the $\cN=1$ superconformal symmetry generated by
$\tT(z)$, $\tG^0(z)$. In other words, $\tG^a(z)$ 
behaves as spin 3/2 primary fields with respect to 
the deformed stress tensor $\tT(z)$;
\begin{eqnarray}
 \tT(z) \tG^a(w) \sim \frac{\frac{3}{2}}{(z-w)^2} \tG^a(w)
 + \frac{1}{z-w} \partial \tG^a(w)
\end{eqnarray}
The central charge is shifted as 
\begin{eqnarray}
 \tc= c+ 3\cQ^2 \equiv 
6 \left(1 -\frac{1}{N} \right)
+ 3 \cQ^2.
 \label{tc}
\end{eqnarray}

In the end, we obtain the modified large $\cN=4$ SCA 
generated by
$\{\tT,\, \tG,\, A^{\pm,i}, \, Q^a, \,  U\}$  
as follows;
\begin{eqnarray}
 && \tG^a(z)\tG^b(w) \sim 
\frac{2\tc}{3}\frac{\delta^{ab}}{(z-w)^3}
  + \frac{8i}{(z-w)^2}\left\{\tgamma \al^{+,i}_{ab}A^{+,i}(w)
 + (1-\tgamma) \al^{-,i}_{ab}A^{-,i}(w)\right\} \nn
&& \hspace{1cm}
 + \frac{4i}{z-w}\left\{\tgamma \al^{+,i}_{ab}\partial
 A^{+,i}(w)
 + (1-\tgamma) \al^{-,i}_{ab}\partial A^{-,i}(w)\right\}
 + \frac{2 \delta^{ab}}{z-w} \tT(w)~, \nn
&& A^{\pm,i}(z)A^{\pm,j}(w) \sim 
\frac{\frac{k^{\pm}}{2}\delta^{ij}}{(z-w)^2} 
+ \frac{i \ep^{ijk}}{z-w} A^{\pm,k}(w) ~, \nn
&& Q^a(z)Q^b(w) \sim \frac{\frac{k^++k^-}{2}}{z-w} \delta^{ab}~,
~~~ U(z)U(w) \sim \frac{\frac{k^++k^-}{2}}{(z-w)^2}~, \nn
&& A^{\pm,i}(z)\tG^a(0) \sim 
\left(\tgamma - \frac{1}{2}\mp \frac{1}{2}\right)
\frac{2 \al^{\pm,i}_{ab}}{(z-w)^2}Q^b(w) + 
\frac{i\al^{\pm,i}_{ab}}{(z-w)}\tG^b(w)~, \nn
&& A^{\pm,i}(z)Q^a(w) \sim \frac{i\al^{\pm,i}_{ab}}{z-w}Q^b(w)
~, \nn
 && \tT(z) U(w) \sim 
\sqrt{\frac{k+2}{2}}\frac{i\cQ}{(z-w)^3}
 + \frac{1}{(z-w)^2} U(w) + \frac{1}{(z-w)}\partial U(w)~, \nn
&& Q^a(z)\tG^b(w) \sim 
\sqrt{\frac{k+2}{2}}
\frac{i\cQ \delta_{ab}}{(z-w)^2} +  
\frac{2}{z-w} \left\{
 \al^{+,i}_{ab}A^{+,i}(w) - \al^{-,i}_{ab}A^{-,i}(w)
\right\} + \frac{\delta^{ab}}{(z-w)} U(w) ~, \nn
&& U(z)\tG^b(w) \sim \frac{1}{(z-w)^2} Q^a(w)~,
\label{large N=4 deformed}
\end{eqnarray}
where $\tgamma$ is defined as 
\begin{eqnarray}
 \tgamma := \gamma - \frac{\cQ}{\sqrt{2(k+2)}} \equiv \frac{1}{N} - \frac{\cQ}{\sqrt{2N}} .
\label{tgamma}
\end{eqnarray}

Here we point out a fact that will be crucial in our arguments:
even though the central charge \eqn{tc} depends on the background charge $\cQ$, 
the {\em effective central charge\/} \cite{KutS-ceff}, which counts the net degrees of freedom, 
is unchanged under the deformation by linear dilaton. 
In the relevant system, the effective central charge should be 
\begin{equation}
c_{\msc{eff}} = c \equiv 6 \left(1- \frac{1}{N}\right),
\label{c eff}
\end{equation}
irrespective of the value of $\cQ$.


~


To conclude this section, we note that 
the whole SCA \eqn{large N=4 deformed}
reduces to the small $\cN=4$ SCA, if we choose particular values of $\cQ$.
In fact, inspecting the OPE of $A^{\pm,i}(z)\tG^a(0)$ given in \eqn{large N=4 deformed}, 
we find that spin 1/2 currents $Q^a$ decouple when we set $\tgamma=0$ or $\tgamma=1$.
We thus obtain the next two `small $\cN=4$ points':

\begin{description}

\item[case 1. $\cQ= \sqrt{\frac{2}{N}}$ : small $\cN=4$ SCA of level $k^-=1$]

~

This case is just the familiar CHS superconformal 
system \cite{CHS}. We have $\tc=6$ and $\tgamma = 0$.
Then, $A^{+,i}(z)$, $Q^a(z)$, $U(z)$ are decoupled,
and  
$\{\tT(z),\, \tG^a(z),\, A^{-,i}(z)\}$ generate the small $\cN=4$ SCA of level 1.


\item[case 2. $\cQ=-(N-1)\sqrt{\frac{2}{N}}$ : small $\cN=4$
SCA of level $k^+=N-1$]

~

In this case, we have $\tc= 6(N-1)$ and $\tgamma= 1$.  
Then, $A^{-,i}(z)$, $Q^a(z)$, $U(z)$ are decoupled,
and $\{\tT(z),\,\tG^a(z),\, A^{+,i}(z)\}$ 
generate the small $\cN=4$ SCA with level $k^+=N-1$.

\end{description}

~

These types of reductions from the large $\cN=4$ to the small
$\cN=4$ with level $k^+$ or $k^-$ have been already discussed in 
\cite{STV,Ivanov:1988rt}, and also potentially utilized in \cite{Matsuda} in order
to construct the Feigin-Fuchs representation of $\cN=4$ SCFT.
In the context of string theory on the NS5-NS1 background, 
the case 1 
is identified with the world-sheet
CFT for the `short string' sector (or that describing the `Coulomb branch
tube'), while the case 2 
corresponds to the `long string' sector (or the `Higgs branch
tube') \cite{SW}. Therefore, they are expected to be dual to each other from the viewpoints of 
$\mbox{AdS}_3/\mbox{CFT}_2$-duality.
In our discussions later, this fact would suggest the existence of two different descriptions 
of the umbral moonshine \cite{umbral1,umbral2} based on the case 1 and 2.
Indeed, these two theories have the equal effective central charge as was already mentioned, 
which implies the essentially same asymptotic growth 
of massive (non-BPS) excitations 
characterizing the moonshine phenomena.

~


\section{Elliptic Genera of $\cN=4$ Liouville Model}
\label{EG N=4 Liouville}

\subsection{Sketch of Outline}
\label{outline}

Our main purpose is  to evaluate the elliptic genera of 
the $\cN=4$ Liouville theory
with  suitable Liouville potentials.
However, it seems hard to directly carry out this computation because of 
the complexity of the $\cN=4$ Liouville potentials. 

We shall thus take another route: regard the relevant $\cN=4 $ superconformal system 
as the $\bz_N$-orbifold of 
$$
SU(2)_N/U(1) \otimes SL(2)_N/U(1) \cong SU(2)_N/U(1) \otimes [\mbox{$\cN=2$ Liouville}]_{\cQ = \sqrt{\frac{2}{N}}},
$$
for the `case 1' ($\hc =2$) \cite{OV}, and then, try to deform the system into the `case 2' ($\hc=2(N-1)$).
The next statements are crucial in our evaluation of the elliptic genus of the case 2.


\begin{description}
\item[(i)] The case 1 and 2 correspond to theories with different central charges. 
Nevertheless, these two theories should possess the equal effective central charge
\eqn{c eff};
$$
c_{\msc{eff}} = 6 \left(1 - \frac{1}{N}\right),
$$
as we already emphasized.
$c_{\msc{eff}}$ characterizes the asymptotic growth of degeneracy of states due to Cardy formula.
In terms of the elliptic genus, since $\cZ^{(\sNS)}(\tau) 
\equiv \cZ^{(\sNS)}(\tau, 0)
:=  q^{\frac{\hc}{8}}  \cZ\left(\tau,  \frac{\tau+1}{2} \right)$ is $S$-invariant, 
\eqn{c eff} implies that the IR-behavior of $\cZ^{(\sNS)}(\tau)$ becomes\footnote
   {Here, we may have a subtlety, since $\cZ^{(\sNS)}(\tau)$ could be non-holomorphic 
due to the existence of modular completions, of which correction terms show a continuous spectrum.
Thus, the Cardy-type argument in this context truly means that the asymptotic growth of the coefficients of $q$-expansion of 
the {\em  holomorphic part} $\cZ^{(\sNS)}_{\msc{hol}}(\tau)$, which only includes a discrete spectrum, is governed by the IR-behavior of 
{\em total}  $\cZ^{(\sNS)}(\tau)$.
}
\begin{equation}
\lim_{\tau_2 \,\rightarrow\, +\infty}\, e^{- 2\pi \tau_2 \frac{c_{\msc{eff}}}{24}}
\, \left|\cZ^{(\sNS)}(\tau)\right| 
\equiv \lim_{\tau_2 \,\rightarrow\, +\infty}\, e^{- 2\pi \tau_2 \left(\frac{1}{4} - \frac{1}{4N} \right)}
\, \left|\cZ^{(\sNS)}(\tau) \right|
< \infty.
\label{cZ NS ceff}
\end{equation}
For the case 1, we can  easily confirm that this condition is satisfied.  
We will later discuss how we can refine the constraint \eqn{cZ NS ceff} due to the  $SU(2)_{N-2}$ symmetry
that is a part of $\cN=4$ superconformal symmetry in the case 2.
Indeed, the resultant constraint will play a crucial role to determine the elliptic genus of the case 2.


\item[(ii)] 
When evaluating the elliptic genera, 
the contributing states in these systems are weighted by {\em different\/}
$U(1)$-currents. 
Namely, setting 
\begin{equation}
\psi^{\pm} := \frac{1}{\sqrt{2}} \left(\psi^3 \mp i \psi^0\right),
\hspace{1cm} 
\chi^{\pm}:= \frac{1}{\sqrt{2}} \left(\psi^1 \pm i \psi^2 \right),
\end{equation}
the $U(1)_R$-current for each case is written as follows;
\begin{itemize}
\item {\bf case 1 :} 
\begin{equation}
J^{(\hc=2)} := 2 A^{-, 3 } = i (\psi^3\psi^0 - \psi^1 \psi^2)
= \psi^+ \psi^- + \chi^+ \chi^-.
\label{J case 1}
\end{equation}

\item {\bf case 2 :} 
\begin{equation}
J^{(\hc=2(N-1))} \equiv 2 A^{+, 3 } : = i (-\psi^3\psi^0 - \psi^1 \psi^2) + 2j^3
= - \psi^+ \psi^- + \chi^+ \chi^- + 2j^3.
\label{J case 2}
\end{equation}


\end{itemize}

\item[(iii)] Elliptic genera should be invariant under generic marginal deformations, at least, for the holomorphic part 
that is contributed from the BPS states. 
Moreover, the non-holomorphic part (`holomorphic anomaly')
only includes the continuous spectrum propagating  
in the asymptotic region where the relevant Liouville potentials are negligible.
These facts imply that both of holomorphic and non-holomorphic terms 
of elliptic genus of the case 2
do not depend on the detail of Liouville potential.

\end{description}

~

Based on these considerations, we propose that the elliptic genus of the case 2 
would be uniquely determined  in the following way;
\begin{itemize}
\item We first evaluate the elliptic genus of case 1, that is, $\cZ^{\msc{case 1}}(\tau,z)$.

\item Secondly, we deform the holomorphic anomaly term in 
$\cZ^{\msc{case 1}}(\tau,z)$, taking account of the distinction 
of $U(1)$-currents between \eqn{J case 1} and \eqn{J case 2}.
The expected non-holomorphic term should  be expanded 
in terms of the $\cN=4$ massive characters of $\hc=2(N-1)$.

\item 
Finally, we determine the holomorphic part of the wanted elliptic genus, which is expected 
to be written in terms of the $\cN=4$ massless characters of $\hc=2(N-1)$.
It will be crucial that the possible ambiguity by adding general holomorphic Jacobi forms 
can be removed by examining the IR-behavior of the NS-sector elliptic genus, 
which extends the argument of effective central charge  given above.

\end{itemize}

~


\subsection{Preliminaries}

\subsubsection{Branching Relation for the $\cN=2$ Minimal Model}


As a preparation, 
we start with recalling the coset construction of $\cN=2$ minimal model.

The $SU(2)_k$ character for the 
spin $\ell/2$-integrable representation is given as 
\begin{eqnarray}
\chi^{(k)}_{\ell}(\tau,z) &:=& \frac{1}{i \th_1 (\tau,z)} \left[\Th{\ell+1}{k+2} (\tau,z) - 
\Th{\ell+1}{k+2} (\tau,- z)\right]
\nn
& \equiv & \frac{2}{i \th_1 (\tau,z)} \Theta^{[-]}_{\ell+1, k+2}(\tau,z),
\hspace{1cm} 
(\ell =0, 1, \ldots, k)
\end{eqnarray}
and the string function $c^{\ell}_m(\tau)$  is defined by the expansion;
\begin{equation}
\chi^{(k)}_{\ell}(\tau,z) = \sum_{m\in \bz_{2k}}\, c^{\ell}_m (\tau) \Th{m}{k}(\tau,z).
\end{equation}


The branching relation describing the $\cN=2$ minimal model as the supercoset 
$$
\frac{SU(2)_k \times SO(2)_1}{U(1)_{k+2}},
$$
is written as \footnote
{$s=0,2 \, (1,3)$ describe free fermions in NS (R) sector.}
\begin{eqnarray}
&& \chi_{\ell}^{(k)}(\tau,w)\Th{s}{2}(\tau,w-z)
=\sum_{m\in \bsz_{2(k+2)}}
\chi_m^{\ell,s}(\tau,z)\Th{m}{k+2}(\tau,w-2z/(k+2)), 
\label{branching} 
\end{eqnarray}
where the branching function $\chi_m^{\ell,s}(\tau,z)$ is explicitly written as  
\begin{equation}
\chi_m^{\ell,s}(\tau,z)=\sum_{r\in \bsz_k}c^{\ell}_{m-s+4r}(\tau)
\Th{2m+(k+2)(-s+4r)}{2k(k+2)}(\tau,z/(k+2)).
\label{chi lms}
\end{equation}
The characters of the $\cN=2$ minimal model are written in terms of the branching functions as follows;
\begin{eqnarray}
\ch{(\sNS)}{\ell,m}(\tau,z) & = & \chi^{\ell,0}_m(\tau,z) + \chi^{\ell, 2}_m(\tau,z),
\nn
\ch{(\stNS)}{\ell,m}(\tau,z) & = & \chi^{\ell,0}_m(\tau,z) - \chi^{\ell, 2}_m(\tau,z),
\nn
\ch{(\sR)}{\ell,m}(\tau,z) & = & \chi^{\ell,1}_m(\tau,z) + \chi^{\ell, 3}_m(\tau,z),
\nn
\ch{(\stR)}{\ell,m}(\tau,z) & = & \chi^{\ell,1}_m(\tau,z) - \chi^{\ell, 3}_m(\tau,z).
\label{ch min}
\end{eqnarray}

Now, the parameter $w$ in \eqn{branching} is interpreted as the deformation parameter
$\tgamma$ (or $\cQ$) in the previous section. 
It is explicitly identified as 
\begin{equation}
w= 2 \tgamma z,
\label{rel w tgamma}
\end{equation}
and the corresponding $U(1)_R$-current is given as  
\begin{equation}
J^{\tgamma} = 2 \tgamma (A^{+, 3} - A^{-, 3}) +  2 A^{-, 3}\equiv 
2 \tgamma \left(j^3 - \psi^+ \psi^- \right) + \left(\psi^+ \psi^- + \chi^+ \chi^- \right).
\label{J tgamma}
\end{equation}

The relevant branching relations are summarized as 
\begin{itemize}
\item $\tgamma =0$ {\bf (case 1) :}
\begin{equation}
\chi_{\ell}^{(N-2)}(\tau,0)\Th{s}{2}(\tau,-z)
=\sum_{m\in \bsz_{2N}}
\chi_m^{\ell,s}(\tau,z)\Th{m}{N}\left(\tau,-\frac{2z}{N}\right),
\label{branching 0} 
\end{equation}

\item $\tgamma =1$ {\bf (case 2) :}
\begin{equation}
\chi_{\ell}^{(N-2)}(\tau,2z)\Th{s}{2}(\tau,z)
=\sum_{m\in \bsz_{2N}}
\chi_m^{\ell,s}(\tau,z)\Th{m}{N}\left(\tau,\frac{2(N-1) z}{N} \right). 
\label{branching 2z} 
\end{equation}


\end{itemize}

~


\subsubsection{Modular Completions}

Let us introduce some notations.
For $N (\in \bz_{>0})$, we set
\begin{equation}
f^{(N)}(\tau,z) := \sum_{n\in \bz}\, \frac{y^{2N n} q^{N n^2}}{1-yq^n},
\hspace{1cm} 
\left(q\equiv e^{2\pi i \tau}, ~ y\equiv e^{2\pi i z}\right),
\label{Appell}
\end{equation}
and for $N, K (\in \bz_{>0})$, 
\begin{eqnarray}
F^{(N,K)}(v,a;\tau,z) & := & \frac{1}{N} \sum_{b\in \bz_N}\, e^{-2\pi i \frac{vb}{N}} y^{\frac{2Ka}{N}} q^{\frac{K a^2}{N}}\,
f^{(NK)}\left(\tau, \frac{z+a\tau+b}{N}\right)
\nn
& \equiv & \sum_{n\in a + N \bz}\, \frac{(y q^n)^{\frac{v}{N}} y^{\frac{2K n}{N}} q^{\frac{K n^2}{N}}}{1-yq^n}.
\label{F}
\end{eqnarray}
We often use the abbreviation $F^{(N)}(v,a) \equiv F^{(N,1)}(v,a)$.
Note that the extended discrete character of the $SL(2)/U(1)$-supercoset with
$\hc = 1+ \frac{2K}{N}$ \cite{ES-L,ES-NH}
is written as 
\begin{equation}
\chid^{(N,K)}(v,a;\tau,z) = \frac{\th_1(\tau,z)}{i\eta(\tau)^3} \, F^{(N, K)}(v,a;\tau,z).
\end{equation}

The modular completion of $f^{(N)}(\tau,z)$ is defined as \cite{Zwegers};
\begin{equation}
\hf^{(N)}(\tau,z) := f^{(N)}(\tau,z) - \frac{1}{2} \sum_{m\in \bz_{2N}}\, R_{m,N}(\tau) \Th{m}{N}\left(\tau, 2z \right),
\label{hAppell} 
\end{equation}
where we set
\begin{eqnarray}
R_{m,N}(\tau) &:=  & \sum_{\nu \in m+2N\bz}\, \sgn(\nu + 0) 
\erfc\left(\sqrt{\frac{\pi \tau_2}{N}} \left|\nu\right|\right)\, 
q^{- \frac{\nu^2}{4N}}
\nn
&\equiv & \frac{1}{i\pi}\, \sum_{\nu \in m+2N\bz}\,
\int_{\br- i0} dp \, \frac{e^{-\pi \tau_2 \frac{p^2+\nu^2}{N}} }{p-i\nu}\,
q^{- \frac{\nu^2}{4N }}.
\label{RmN}
\end{eqnarray}
Here, we denote $\tau_2 \equiv \mbox{Im}\, \tau$ and $\erfc(x)$ is the error function \eqn{Erfc}.
$\hf^{(N)}(\tau,z)$ is a weight $1$, index $N$ (real analytic) weak Jacobi form \cite{Zwegers}.

It is useful to rewrite \eqn{hAppell} by introducing the anti-symmetrization 
$$
f^{(N)\, [-]}(\tau,z) :=  \frac{1}{2} \left[ 
f^{(N)}(\tau,z) - f^{(N)}(\tau,-z)
\right] .
$$
We note 
$$
\hf^{(N)\, [-]}(\tau,z) :=  \frac{1}{2} \left[ 
\hf^{(N)}(\tau,z) - \hf^{(N)}(\tau,-z)
\right] 
\equiv \hf^{(N)}(\tau,z),
$$
since the completion $\hf^{(N)}(\tau,z)$ gives an odd function of $z$ (see {\em e.g.} \cite{ES-NH}).
Then, we obtain 
\begin{eqnarray}
\hf^{(N)}(\tau,z) & = & f^{(N)\,[-]}(\tau,z) - 
\frac{1}{2} \sum_{m\in \bz_{2N}}\, R_{m,N}(\tau) \Th{m}{N}^{[-]}\left(\tau, 2z \right)
\nn
& = &  f^{(N)\,[-]}(\tau,z) -  \sum_{m=1}^{N-1}\, R_{m,N}(\tau) \Th{m}{N}^{[-]}\left(\tau, 2z \right).
\label{hAppell 2}
\end{eqnarray}
In the second line of \eqn{hAppell 2}, we made use of the facts that 
$\Th{Nj}{N}^{[-]}(\tau,z) \equiv 0$ $(\any j \in \bz)$ and  
$$
R_{-m,N}(\tau) = - R_{m,N}(\tau) + 2 \delta^{(2N)}_{m,0}, \hspace{1cm} R_{m+2N, N}(\tau) = R_{m,N}(\tau). 
$$
%
We also introduce the modular completion of $F^{(N,K)}(v,a)$ \eqn{F} \cite{ES-NH};
\begin{eqnarray}
\hspace{-1cm}
\hF^{(N,K)}(v,a;\tau,z) & := & \frac{1}{N} \sum_{b\in \bz_N}\,
 e^{-2\pi i \frac{vb}{N}} y^{\frac{2Ka}{N}} q^{\frac{K a^2}{N}}\,
\hf^{(NK)}\left(\tau, \frac{z+a\tau+b}{N}\right)
\nn
& \equiv & F^{(N,K)}(v,a;\tau,z) - \frac{1}{2} \sum_{j\in \bz_{2K}}\, R_{v+Nj, NK}(\tau) 
\Th{v+Nj+2Ka}{NK} \left(\tau, \frac{2z}{N}\right).
\label{hF NK}
\end{eqnarray}
Especially, for the cases of $K=1$, the function
$
\hF^{(N)}(v,a) \equiv \hF^{(N,1)}(v,a)
$
becomes
\begin{eqnarray}
&& \hF^{(N)}(v,a;\tau,z) 
= F^{(N)}(v,a;\tau,z) - \frac{1}{2} \sum_{j\in \bz_2}\, R_{v+Nj, N}(\tau) 
\Th{v+Nj+2a}{N} \left(\tau, \frac{2z}{N}\right).
\label{hF N}
\end{eqnarray}


~


\subsection{Case 1 models}

We propose the following elliptic genus for the case 1 models
with $\hc=2$,
\begin{eqnarray}
\cZ^{\msc{case 1}}(\tau,z) & = & \frac{1}{N} \sum_{a,b \in \bz_N} \,
\cZ^{\msc{(min)}}_{[a,b]}(\tau,z) \, \cZ^{SL(2)/U(1)}_{[a,b]}(\tau,z)
\nn
& = & 
-\frac{\th_1(\tau,z)}{i\eta(\tau)^3} \,
\sum_{\ell=0}^{N-2} \, \sum_{a\in \bz_N} \, \ch{(\stR)}{\ell, \ell+1+2a} (\tau,z) \,
\hF^{(N)}(\ell+1,  a ;\tau, - z),
\label{cZ ALE} 
\end{eqnarray}
where we set
\begin{eqnarray}
\cZ^{\msc{(min)}}_{[a,b]}(\tau,z) &:=& 
(-1)^{a+b+ab} \, q^{\frac{N-2}{2N} a^2}  y^{\frac{N-2}{N} a} e^{i\pi \frac{N-2}{N} ab}\,
\cZ^{\msc{(min)}} (\tau,z+a\tau+b),
\label{cZ min ab}
\\
\cZ^{\msc{(min)}} (\tau,z) & : = & \frac{\th_1\left(\tau, \frac{N-1}{N}z \right)}{\th_1 \left(\tau, \frac{z}{N} \right)} 
\equiv \sum_{\ell=0}^{N-2} \, \ch{(\stR)}{\ell,\ell+1} (\tau,z),
\label{cZ min}
\end{eqnarray}
for the minimal model ($SU(2)/U(1)$-sector) \cite{Witten-E2}, and 
\begin{eqnarray}
\cZ^{SL(2)/U(1)}_{[a,b]}(\tau,z) & := & (-1)^{a+b+ab} \, q^{\frac{N+2}{2N} a^2}  
y^{\frac{N+2}{N} a} e^{i\pi \frac{N+2}{N} ab}\,
\nn
&& \hspace{1cm} 
\times \cZ^{SL(2)/U(1)}(\tau,z+a\tau+b),
\label{cZ SL(2)/U(1) ab}
\\
\cZ^{SL(2)/U(1)}(\tau,z) & : = & \frac{\th_1(\tau,z)}{i\eta(\tau)^3} \, \hf^{(N)}\left(\tau, \frac{z}{N}\right)
\nn
& \equiv & \frac{\th_1(\tau,z)}{i\eta(\tau)^3} \, \sum_{v\in \bz_N} \, \hF^{(N)}(v,0;\tau,z),
\label{cZ SL(2)/U(1)}
\end{eqnarray}
for the $SL(2)/U(1)$-sector \cite{Troost,ES-NH}.


The relevant branching relation is given by \eqn{branching 0}, namely,
\begin{eqnarray}
&&
\sum_{m\in \bz_{2N}}\, \ch{(\stR)}{\ell,m}(\tau,z) \Th{m}{N}\left(\tau , - \frac{2z}{N}\right) = 
-i \th_1(\tau,z) \chi^{(k)}_{\ell} (\tau, 0)
\nn
&& \hspace{2cm}
= - \frac{\th_1(\tau,z)}{i \pi } \oint_{w=0} \frac{dw}{w} \, \frac{\Th{\ell+1}{N}^{[-]}(\tau, 2w) }{\th_1(\tau,2w)} .
\label{branching ALE}
\end{eqnarray}
By using this identity, we can show 
\begin{eqnarray}
&& [\mbox{non-hol. part of $\cZ^{\msc{case 1}}$}]
\nn
&& 
\hspace{1cm}
 = \frac{1}{2}  \frac{\th_1(\tau,z)}{i\eta(\tau)^3} \, 
\sum_{\ell=0}^{N-2}\, \sum_{a\in \bz_N}\,\sum_{j\in \bz_2}\, 
R_{\ell+1 +Nj,N} \Th{\ell+1 +Nj+2a}{N}\left(-\frac{2z}{N}\right)\,
\ch{(\stR)}{\ell, \ell+1+2a}(z)
\nn
&& 
\hspace{1cm}
 = - \frac{\th_1(\tau,z)^2}{\eta(\tau)^3} 
\frac{1}{i \pi } \oint_{w=0} \frac{dw}{w} \, 
\frac{1 }{i \th_1(\tau,2w)} \sum_{\ell=0}^{N-2} \, 
R_{\ell+1, N} \Th{\ell+1}{N}^{[-]}(\tau, 2w).
\label{eval shadow ALE}
\end{eqnarray}
Therefore, recalling \eqn{hAppell 2}, 
we find that 
\begin{eqnarray}
&& \frac{\partial}{\partial \bar{\tau}} \cZ^{\msc{case 1}}(\tau)
\nn
&& 
\hspace{1cm}
= \frac{\partial}{\partial \bar{\tau}} \left[  \frac{\th_1(\tau,z)^2}{\eta(\tau)^3} 
\frac{1}{i \pi } \oint_{w=0} \frac{dw}{w} \, 
\frac{\hf^{(N)}(\tau,w) }{i \th_1(\tau,2w)} \right]
\nn
&& 
\hspace{1cm}
= \frac{\partial}{\partial \bar{\tau}} \left[  \frac{\th_1(\tau,z)^2}{\eta(\tau)^3} 
\frac{1}{i \pi } \oint_{w=0} \frac{dw}{w} \, 
\frac{\hf^{(N)}(\tau,w) }{i \th_1(\tau,2w)} e^{(N-2) G_2(\tau) w^2}\right],
\label{eval shadow ALE 2}
\end{eqnarray}
where $G_2(\tau)$ is the (unnormalized) 2nd Eisenstein series \eqn{def G2}.  
In order to gain the third line of \eqn{eval shadow ALE 2}, 
we made use of the fact that the function
$$
\frac{\partial}{\partial \bar{\tau}} \frac{\hf^{(N)}(\tau,w) }{i \th_1(\tau,2w)}
\equiv \frac{\partial}{\partial \bar{\tau}} \frac{\hf^{(N), [-]}(\tau,w) }
{i \th_1(\tau,2w)},
$$
is holomorphic with respect to $w$.
It is important that the integrand 
\begin{equation}
g^{(N)}(\tau,w) := 
\frac{\hf^{(N)}(\tau,w) }{i \th_1(\tau,2w)} e^{(N-2) G_2(\tau) w^2},
\label{gN}
\end{equation}
possesses the correct modular property due to the factor $e^{(N-2) G_2(\tau) w^2}$
(see the formula \eqn{G2 S}),  that is, 
\begin{equation}
g^{(N)}\left(-\frac{1}{\tau}, \frac{w}{\tau} \right) = - \sqrt{-i\tau}\, g^{(N)}(\tau,w).
\label{S g}
\end{equation}
Thus, we conclude that 
\begin{eqnarray}
\cZ^{\msc{case 1}}(\tau, z) & = & [\mbox{holomorpchi Jacobi form}]
\nn
&& ~~~
+ \frac{\th_1(\tau,z)^2}{\eta(\tau)^3} 
\frac{1}{i \pi } \oint_{w=0} \frac{dw}{w} \, 
\frac{\hf^{(N)}(\tau,w) }{i \th_1(\tau,2w)} e^{(N-2) G_2(\tau) w^2}.
\end{eqnarray}
It is easy to identify the first term because of the uniqueness of 
weak  Jacobi form of weight 0, index 1, that is, 
\begin{equation}
\phi_{0,1}(\tau,z) \equiv \frac{1}{2} \cZ^{\msc{K3}}(\tau,z) \equiv 
4 \left[\left(\frac{\th_3(\tau,z)}{\th_3(\tau,0)}\right)^2
+ \left(\frac{\th_4(\tau,z)}{\th_4(\tau,0)}\right)^2 + \left(\frac{\th_2(\tau,z)}{\th_2(\tau,0)}\right)^2
\right].
\end{equation}
We also recall that the Witten index should be 
\begin{equation}
\cZ^{\msc{case 1}}(\tau, 0) = N-1,
\label{WI case 1}
\end{equation}
(see {\em e.g.} \cite{ES-BH}), 
which fixes the normalization of the holomorphic term. 

In this way, we finally achieve a simple formula;
\begin{eqnarray}
\hspace{-1cm}
\cZ^{\msc{case 1}}(\tau, z) & = & \frac{N-1}{12} \phi_{0,1}(\tau,z)
+ \frac{\th_1(\tau,z)^2}{\eta(\tau)^3} 
\frac{1}{i \pi } \oint_{w=0} \frac{dw}{w} \, 
\frac{\hf^{(N)}(\tau,w) }{i \th_1(\tau,2w)} e^{(N-2) G_2(\tau) w^2}.
\label{formula cZ ALE}
\end{eqnarray}

~


\subsubsection{Useful facts on the formula \eqn{formula cZ ALE} }
\label{remark cZ ALE}

We exhibit some useful computations related to the resultant formula \eqn{formula cZ ALE} and
make a few remarks.
For convenience, 
we define the non-holomorphic modular form $\hH^{(N)}(\tau)$ of weight 2  by the relation;
\begin{eqnarray}
&& \cZ^{\msc{case 1}}(\tau, z)  =  \frac{N-1}{12} \phi_{0,1}(\tau,z)
+ \frac{\th_1(\tau,z)^2}{\eta(\tau)^6} \, \hH^{(N)}(\tau).
\label{def hH N}
\end{eqnarray}
Namely, we set
\begin{equation}
\hH^{(N)} (\tau) := 
\frac{\eta(\tau)^3}{i \pi } \oint_{w=0} \frac{dw}{w} \, 
\frac{\hf^{(N)}(\tau,w) }{i \th_1(\tau,2w)} e^{(N-2) G_2(\tau) w^2}.
\label{def hHN}
\end{equation}
Then, by substituting the `Poincare series' formula \cite{ES-nhP};
\begin{equation}
\hf^{(N)}(\tau,z) = \frac{i}{2\pi} \sum_{\la \in \La} \, 
\frac{e^{-\frac{\pi }{\tau_2} N  \left\{\left|\la \right|^2 + 2\bar{\la} z + z^2 \right\}}}{\la+z},
\hspace{1cm} 
\left(\La \equiv \bz \tau +\bz \right),
\label{formula nhP 1}
\end{equation}
into \eqn{def hHN},
we can rewrite it as 
\begin{eqnarray}
\hH^{(N)}(\tau) & = & 
\frac{1}{4\pi^2} \left[ N \widehat{G}_2(\tau) + \frac{\del}{\del w} \sum_{\la \in \La'}\, 
\left.
\frac{e^{-\frac{\pi}{\tau_2} N \left\{\left|\la \right|^2 + 2\bar{\la} w + w^2 \right\}}}{\la+w}
\right|_{w=0}  \right]
\nn
&=& \frac{1}{4\pi^2} \left[ N \hG_2(\tau) -  
\sum_{\la\in \La'} \, \frac{e^{- \frac{\pi}{\tau_2} N \left|\la\right|^2 }}{\la^2} 
\left\{
1+ \frac{2 \pi N}{\tau_2} \left|\la\right|^2
\right\} \right].
\label{formula hHN}
\end{eqnarray}
Here, the summation is taken over $\la \in \La' \equiv \bz \tau + \bz - \{0\}$, and 
$
\dsp
\hG_2(\tau) \equiv G_2(\tau) - \frac{\pi}{\tau_2} 
$
denotes the modular completion of  (unnormalized) 2nd Eisenstein series \eqn{def hG2}. 
We present an explicit derivation of \eqn{formula hHN} in Appendix B.


The above result 
\eqn{formula hHN}  suggests that the holomorphic part $H^{(N)}(\tau)$
would be 
\begin{eqnarray}
&& H^{(N)}(\tau)
\sim
\frac{N}{4\pi^2} G_2(\tau) 
+ \frac{1}{4\pi^2} \frac{\del}{\del w} \sum_{\la = m\tau +n \in \La'}\, 
\left.
\frac{q^{N m^2} e^{2\pi i (2N) m w}}{\la+w}
\right|_{w=0} .
\label{guess HN} 
\end{eqnarray}
However, the double series appearing in \eqn{guess HN} does not converge, 
and thus we have to be more careful. 
To this end, 
we introduce the symbol of the `principal value';
\begin{equation}
{\sum_{n\neq 0}}^P \, a_n := \lim_{N\,\rightarrow\, \infty} \sum_{n=1}^N \, \left(a_n + a_{-n}\right), 
\hspace{1cm} 
{\sum_{n\in \bz}}^P \, a_n := a_0 + {\sum_{n\neq 0}}^P \, a_n,
\label{def sumP}
\end{equation}
and the correct expression of $H^{(N)}(\tau)$ should be
\begin{eqnarray}
\hspace{-5mm}
&& H^{(N)}(\tau)
=
\frac{N}{4\pi^2} G_2(\tau) \left.
+ \frac{1}{4\pi^2} \frac{\del}{\del w} \left[ \sum_{m \neq 0}\, {\sum_{n\in \bz}}^P\, 
\frac{q^{N m^2} e^{2\pi i (2N) m w}}{\la+w}  
+ {\sum_{n \neq 0}}^P \, \frac{1}{w+n}
\right] \right|_{w=0}.
\label{formula HN 1}
\end{eqnarray}
A rigorous derivation of \eqn{formula HN 1} is again presented in Appendix B.


One would be interested in the $q$-expansion of $H^{(N)}(\tau)$.
Substituting the familiar formula \eqn{def G2} as well as 
\begin{eqnarray*}
&& 
\sum_{n=1}^{\infty}\, \frac{1}{n^2} \equiv \zeta(2) = \frac{\pi^2}{6},
\hspace{1cm}
\frac{i}{2\pi} \, {\sum_{n\in \bz}}^P \,\frac{1}{z+n} = \frac{1}{2} + \frac{y}{1-y} = -\frac{1}{2} + \frac{1}{1-y},
\end{eqnarray*}
into \eqn{formula HN 1},  we  readily obtain 
\begin{eqnarray}
&& H^{(N)}(\tau) = 
\frac{N-1}{12} - 2N \sum_{n=1}^{\infty} \, \frac{n q^n}{1-q^n}
+  2 \sum_{m=1}^{\infty} \, q^{Nm^2}  \left[ \frac{q^{m}}{\left(1 -q^m \right)^2}
+  Nm \frac{1+q^m}{1-q^m} \right].
\label{q-exp HN}
\end{eqnarray}


Let us  remark:
\begin{itemize}
\item The constant term $ \frac{N-1}{12}$ is  anticipated. It precisely cancels 
the `graviton term' included in the holomorphic Jacobi form $ \frac{N-1}{12} \phi_{0,1}(\tau,z) $
appearing in  $\cZ^{\msc{case 1}}(\tau,z) $ \eqn{formula cZ ALE}.

In fact,  the term 
$
\left(\frac{\th_2(\tau, z)}{\th_2(\tau)}\right)^2 
$
in $\phi_{0,1}(\tau,z)$ yields the leading contribution (`graviton term'), after making the spectral 
flow $z\, \mapsto \, \frac{\tau+1}{2}$. 
We thus  obtain the evaluation 
\begin{equation}
[\mbox{graviton term}]
\sim - \frac{N-1}{3} \left( \frac{\th_4(\tau,z)}{\th_2(\tau)}\right)^2
\sim - \frac{N-1}{12} q^{- \frac{1}{4}},
\hspace{1cm} (\tau \, \rightarrow \, i\infty),
\label{formula cZ NS IR eval 1}
\end{equation}
by using
$
q^{\frac{1}{8}} y^{\frac{1}{2}}\, \th_2\left(\tau, z+ \frac{\tau+1}{2}\right) 
= -i \th_4(\tau,z).
$
On the other hand, the first term of \eqn{q-exp HN} yields 
\begin{equation}
\frac{N-1}{12} \frac{\th_3(\tau,z)^2}{\eta(\tau)^6} \sim \frac{N-1}{12} q^{-\frac{1}{4}}, \hspace{1cm} (\tau \, \rightarrow \, i\infty).
\label{formula cZ NS IR eval 2}
\end{equation}


\item 
It is easily confirmed that 
the function $\frac{12}{N-1} H^{(N)}(\tau) $
has the $q$-expansion such as
\begin{equation}
\frac{12}{N-1} H^{(N)}(\tau) = 1 - \sum_{n=1}^{\infty}\, a_n q^n,
\end{equation}
with integer coefficients $a_n$ as long as 
$N-1$ divides 24.
Moreover, since the second term in \eqn{q-exp HN} looks more dominant than the third one,
we expect that all the coefficients $a_n$ are positive.

\end{itemize}


~

Amusingly, for the special case $N=2$ of Mathieu moonshine, 
we have an alternative expression for the $\hH^{(2)}(\tau)$ as
\begin{eqnarray}
\hH^{(2)} (\tau) &=& 
\frac{\eta(\tau)^3}{2\pi i } \oint_{w=0} \frac{dw}{w} \, 
\frac{\htf^{(1/2)}(\tau,w) }{i \th_1(\tau,w)} 
\nn
& = & \frac{1}{4\pi^2}
\left[G_2(\tau) +
 \sum_{\la = m\tau+n \in \La'} \, (-1)^{m+n+mn}\,
\frac{ 2\pi i m e^{- \frac{\pi}{2\tau_2} \left|\la\right|^2 }}{\la}  
\right].
\label{formula hH2}
\end{eqnarray}
The first line follows from the identity\footnote
   {The holomorphic part of the identity \eqn{id hf htf} essentially means 
the familiar equivalence between the $\cN=4$ massless character of level 1
and the spectral flow sum of the $\cN=2$ massless matter characters with $\hc=2$ \cite{ET}, which is also presented in \cite{CH}.
The identity \eqn{id hf htf} claims that this equivalence still holds {\em after taking the modular completions\/}, 
and it is surely a non-trivial identity. One of its proof is obtained by setting $N=2$ in the identity \eqn{main id hf hF} which 
we will prove. Note that the minimal character $\ch{(\stR)}{0,m}(\tau,z)$ just reduces to 
the mod 2 Kronecker delta : $\delta^{(2)}_{m,1} - \delta^{(2)}_{m,-1}$ in the case of $N=2$ ($\hc_{\msc{min}}=0$).} 
\begin{equation}
\hf^{(2)}(\tau,z) = \frac{\th_1(\tau, 2z)}{2 \th_1(\tau,z)} \, \htf^{(1/2)}(\tau,z),
\label{id hf htf}
\end{equation}
where we set 
\begin{eqnarray}
\htf^{(1/2)}(\tau,z) &:=& \hF^{(2)}(1,0;\tau,z) - \hF^{(2)}(1,1;\tau,z)
\\
& \equiv &  \tilde{f}^{(1/2)} (\tau,z) + [\mbox{non-hol. correction}],
\nn
\tilde{f}^{(1/2)} (\tau,z) 
& :=  & \sum_{n \in \bz}\, \frac{(-1)^n (yq^n)^{\frac{1}{2}}}{1-y q^n} y^{n} q^{\frac{1}{2} n^2} .
\end{eqnarray}
The second line of \eqn{formula hH2}\footnote
   {This is equivalent to the identity that Zagier gave in his lecture at the Durham workshop, August 2015.  }  is obtained by 
substituting the another formula of the non-holomorphic Poincare series;
\begin{equation}
\htf^{(1/2)}(\tau,z) = \frac{i}{2\pi} \sum_{\la=m\tau+n \in \La} \, (-1)^{m+n+mn}\,
\frac{e^{-\frac{\pi }{2 \tau_2}   \left\{\left|\la \right|^2 + 2\bar{\la} z + z^2 \right\}}}{\la+z}.
\label{formula nhP 2}
\end{equation}
The second line of 
\eqn{formula hH2} is again derived in Appendix B.


We can also rewrite the first line of \eqn{formula hH2} by replacing 
$1/w$  in the integrand 
with an elliptic function $\frac{1}{3} \xi(\tau,w)$ defined by 
\begin{equation}
\xi(\tau,w) := \frac{\del}{\del w} \ln \left(\frac{\th_1(\tau,w)^3}{\th_2(\tau,w) \th_3(\tau,w) \th_4(\tau,w)}\right).
\label{def xi} 
\end{equation}
It is easy to see that $\xi(\tau,z)$ is an elliptic function of order 4 which possesses 
simple poles at $w=0, \, \frac{1}{2}, \, \frac{\tau}{2}, \, \frac{\tau+1}{2}$ with the residues 
$$
\mbox{Res}_{w=0} [\xi(\tau,w)] = 3, ~~~ \mbox{Res}_{w=\frac{1}{2}} [\xi(\tau,w)] = 
\mbox{Res}_{w=\frac{\tau}{2}} [\xi(\tau,w)] = 
\mbox{Res}_{w=\frac{\tau+1}{2}} [\xi(\tau,w)]=-1.
$$
Then, the integrand in \eqn{formula hH2} becomes an elliptic function with 
a cubic pole $w=0$ and simple poles $w=\frac{1}{2}, \, \frac{\tau}{2}, \, \frac{\tau+1}{2}$.
We thus obtain by the contour deformation
\begin{eqnarray}
\hH^{(2)} (\tau) &=& 
\eta(\tau)^3 \, \mbox{Res}_{w=0} \, \left[ \frac{1}{3} \xi(\tau,w) 
\frac{\htf^{(1/2)}(\tau,w) }{i \th_1(\tau,w)} \right]
\nn
& = & - \frac{1}{3} \eta(\tau)^3\,
\sum_{w_0=\frac{1}{2}, \frac{\tau}{2}, \frac{\tau+1}{2}} \,
\mbox{Res}_{w_0} \left[\xi(\tau, w) \frac{\htf^{(1/2)}(\tau,w) }{i \th_1(\tau,w)} \right]
\nn
& = & 
 \frac{1}{3} \eta(\tau)^3 \,
\sum_{w_0=\frac{1}{2}, \frac{\tau}{2}, \frac{\tau+1}{2}} \, 
\frac{\htf^{(1/2)}(\tau,w_0) }{i \th_1(\tau,w_0)}.
\label{formula hH2 2}
\end{eqnarray}
The holomorphic part of R.H.S of \eqn{formula hH2 2} is identical to 
the known expression \cite{EOT} of  Mathieu moonshine, 
which is derived using the relation among $\cN=4$ character formulas at level 1 \cite{ET,EOTY};
\begin{eqnarray}
\hspace{-5mm}
 \ch{(\stR)}{0} (k=1, \ell=0;\tau,z) 
& = & \frac{\th_1(\tau,z)}{i\eta(\tau)^3}\, \widetilde{f}^{(1/2)} (\tau, z)
\nn
&=& \left(\frac{\th_i(\tau,z)}{\th_i(\tau,0)}\right)^2 + 
\frac{\th_1(\tau,z)^2}{\eta(\tau)^3} \, \frac{\widetilde{f}^{(1/2)}(\tau,w_i)}{i \th_1(\tau,w_i)},
\hspace{1cm} (\any i=2,3,4),
\nn
&& \hspace{2cm} 
\left(w_2 \equiv \frac{1}{2}, ~ w_3 \equiv \frac{\tau+1}{2}, ~ w_4 \equiv \frac{\tau}{2}\right).
\label{formula EOTY}
\end{eqnarray}

~


\subsubsection{Comments on the shadow }

As a consistency check of \eqn{formula hHN}, let us evaluate its `shadow' \cite{Zwegers}. 
After a short calculation, we obtain   
\begin{eqnarray}
&& \sqrt{\tau_2} \frac{\del}{\del \bar{\tau}} \hH^{(N)}(\tau) 
=  \frac{i N}{8\pi \tau_2^{3/2}}\sum_{\la \in \La} \, e^{- \frac{\pi}{\tau_2} N \left|\la\right|^2} 
\left( 1 - \frac{2\pi N}{\tau_2}\left|\la \right|^2\right) .
\label{shadow hHN 1}
\end{eqnarray}
On the other hand, due to the Poisson resummation, we find 
\begin{eqnarray}
&& \sum_{r \in \bz_{2N}}\, \Th{r}{N}(\tau,z_L) \overline{\Th{r}{N}(\tau, z_R)}
= \sqrt{\frac{N}{\tau_2}} e^{- \frac{\pi N}{4\tau_2} (z_L- \overline{z_R} )^2 } \,
\sum_{\la \in \La} \,
e^{- \frac{\pi N}{\tau_2} \left\{ \left|\la\right|^2 + \left(\bar{\la} z_L - \la \overline{z_R}\right) \right\} },
\label{theta PR formula}
\end{eqnarray}
and thus, 
\begin{eqnarray}
&& 
\hspace{-1cm}
\sum_{r \in \bz_{2N}}\, \left. \del_{z_L} \Th{r}{N}(\tau,z_L) 
\overline{ \del_{z_R }\Th{r}{N}(\tau, z_R)} \right|_{z_L=z_R=0}
= \frac{\pi N^{3/2}}{2  \tau_2^{3/2}}\sum_{\la \in \La} \, e^{- \frac{\pi}{\tau_2} N \left|\la\right|^2} 
\left( 1 - \frac{2\pi N}{\tau_2}\left|\la \right|^2\right).
\label{theta PR formula 2}
\end{eqnarray}
Therefore, 
introducing the `unary theta function' \cite{Zwegers}
\begin{equation}
S_{r, N}(\tau) :=  \left. \frac{1}{2\pi i} \del_z \Th{r}{N}(\tau,2z)\right|_{z=0}
\equiv \sum_{n\in r+2N \bz}\, n q^{\frac{n^2}{4N}},
\label{def S}
\end{equation}
we finally obtain 
\begin{equation}
\sqrt{\tau_2} \frac{\del}{\del \bar{\tau}} \hH^{(N)} (\tau) = \frac{i}{ 4\sqrt{N}} \sum_{r\in \bz_{2N}} \,
\left|S_{r, N} (\tau) \right|^2.
\label{shadow hHN 2}
\end{equation}
This is the expected result. 
Indeed, by using the familiar property of the `R-function' of \cite{Zwegers};
\begin{equation}
\sqrt{\tau_2} \frac{\del}{\del \bar{\tau}} R_{m, N}(\tau) = - \frac{i}{2 \sqrt{N}} \overline{S_{m,N}(\tau)}.
\label{shadow R}
\end{equation}
Together with \eqn{hAppell 2} and \eqn{def hHN},
we can directly evaluate the shadow of $\hH^{(N)}(\tau)$ as
\begin{eqnarray}
\sqrt{\tau_2} \frac{\del}{\del \bar{\tau}} \hH^{(N)}(\tau) 
& = & - \frac{1}{i\pi} \oint\frac{dw}{w} \, \frac{\eta(\tau)^3}{i \th_1(\tau,2w)}\,
\sum_{v=1}^{N-1}\, \sqrt{\tau_2} \frac{\del}{\del \bar{\tau}} R_{v,N}(\tau) \, \Th{v}{N}^{[-]}(\tau,2w) 
\nn
& = & 
- \frac{1}{2\pi i}
\sum_{v=1}^{N-1} \,
\frac{-i}{2\sqrt{N}}  \overline{S_{v,N}(\tau)}
\left. \del_w \Th{v}{N}(\tau,2w) \right|_{w=0}
\nn
& = &  \frac{i}{4 \sqrt{N}} \sum_{r\in \bz_{2N}} \, \left| S_{r,N}(\tau)\right|^2,
\end{eqnarray}
which coincides with \eqn{shadow hHN 2}.




It may be also useful to evaluate the shadow of $\hf^{(N)}(\tau,z)$ based on the formula of the non-holomorphic Poincare series
\eqn{formula nhP 1};
$$
\hf^{(N)}(\tau,z) = \frac{i}{2\pi} \sum_{\la \in \La}\, \frac{\rho^{(N)}(\la,z)}{\la+z} \equiv 
\frac{i}{2\pi} \sum_{\la \in \La}\, \frac{e^{- \frac{\pi N}{\tau_2} \left\{ \left|\la\right|^2 
+ 2\bar{\la} z + z^2\right\}}}{\la+z},
$$
where we introduced the notation 
\begin{eqnarray}
\rho^{(k)} (\la, z)& := &
e^{- \frac{\pi k}{\tau_2} \left\{\left|\la \right|^2 + 2 \bar{\la} z + z^2\right\}}
\nn
& \equiv & q^{km^2} y^{2km} e^{2\pi i k m n} e^{-\frac{\pi k}{\tau_2} z^2}.
\label{def rho}
\end{eqnarray}
Using
$$
\frac{\del}{\del \bar{\tau}} \rho^{(N)}(\la,z) = \frac{i \pi N}{2 \tau_2^2} \left(\la + z \right)^2 \rho^{(N)}(\la,z),
$$
we easily obtain 
\begin{equation}
\sqrt{\tau_2} \frac{\del}{\del \bar{\tau}} \hf^{(N)}(\tau,z) 
= - \frac{N}{4 \tau_2^{3/2}} \sum_{\la \in \La} (\la+z) \rho^{(N)}(\la,z).
\label{shadow hAppell 1}
\end{equation}
However, 
by differentiating with respect to $\overline{z_R}$ of the both sides of \eqn{theta PR formula},
we find 
\begin{equation}
\sum_{r\in \bz_{2N}} \, \Th{r}{N}(\tau,2z) \overline{S_{r,N}(\tau)} 
= \frac{i  N^{3/2}}{ \tau_2^{3/2}} \sum_{\la \in \La} \, (\la+z) \rho^{(N)}(\la,z).
\end{equation}
We thus obtain 
\begin{equation}
\sqrt{\tau_2} \frac{\del}{\del \bar{\tau}} \hf^{(N)}(\tau,z)  = 
 \frac{i}{4\sqrt{N}} \sum_{r\in \bz_{2N}}\,  \Th{r}{N}(\tau,2z) \overline{S_{r,N}(\tau)},
\label{shadow hf}
\end{equation}
which is consistent with the formulas \eqn{hAppell} and \eqn{shadow R}.

~


\subsection{Elliptic genus of case 2 theory}

Now, let us begin our analysis for the system of `case 2', {\em i.e.}  the $\cN=4$ theory with $\hc=2(N-1)$.
According to our strategy addressed in subsection \ref{outline}, 
we motivate $\cZ^{\msc{case 2}}(\tau,z)$ by modifying the coupling of the variable $z$ in 
$\cZ^{\msc{case 1}}(\tau,z)$. 
The relevant branching relation is now \eqn{branching 2z}, 
in other words, 
\begin{eqnarray}
2 \frac{\th_1(\tau,z)}{\th_1(\tau,2z)} \, \Th{\ell+1}{N}^{[-]} (\tau,2z)
= \sum_{m \in \bz_{2N}}\, \ch{(\stR)}{\ell,m}(\tau,z) \Th{m}{N} \left(\tau, \frac{2(N-1) }{N}z \right).
\label{branching theta}
\end{eqnarray}
This identity can be
interpreted as the $\frac{SU(2)}{U(1)} \times \frac{SL(2)}{U(1)}$-decomposition 
of the $\cN=4$ massive characters with isospin $\frac{j}{2} \equiv \frac{\ell+1}{2}$, conformal weight
$
h \equiv \frac{p^2}{2} + \frac{\ell(\ell+2)}{4N} +\frac{(N-1)^2}{4}+ \frac{1}{4}
$;
\begin{eqnarray}
\hspace{-1cm}
\ch{(\stR)}{}\left(N-1, h, j ;\tau,z\right)
&\equiv& 
(-1)^{N-\ell} q^{\frac{p^2}{2}} \frac{2 \th_1(\tau,z)^2}{i\eta(\tau)^3 \th_1(\tau,2z)} \,
\Th{\ell+1}{N}^{[-]} (\tau,2z)
\nn
&= & (-1)^{N-\ell} q^{\frac{p^2}{2}} \frac{\th_1(\tau,z)}{i\eta(\tau)^3} \, 
\sum_{m \in \bz_{2N}}\, \ch{(\stR)}{\ell,m}(\tau,z) \Th{m}{N} \left(\tau, \frac{2(N-1) }{N}z \right) .
\label{decomp N=4 massive ch}
\end{eqnarray}
This formula suggests that the following elliptic genus of the case 2 theory 
\begin{eqnarray}
\cZ^{\msc{case 2}} (\tau,z) 
&=& \frac{\th_1(\tau,z)}{i\eta(\tau)^3} \, \sum_{\ell=0}^{N-2}\,
\sum_{a \in \bz_{N}}\, \ch{(\stR)}{\ell,\ell+1+2a}(\tau,z) 
\hF^{(N)}(\ell+1, a ; \tau, (N-1)z)
\label{ansatz cZ N=4 L}
\end{eqnarray}
up to an overall phase factor. 
This is obtained by  the replacement: 
$- z \, \mapsto \, (N-1)z$ in the function $\hF^{(N)}(*,*) $ appearing in 
$\cZ^{\msc{case 1}}$ \eqn{cZ ALE}, 
which properly corrects the difference of the $U(1)_R$-charges between the case 1 and 2,
and 
reproduces the expected holomorphic anomaly terms 
expanded by the $\cN=4$ massive characters.
(Recall the branching relations  \eqn{branching 0} and \eqn{branching 2z}.)


%
%


~

\subsubsection{Proof of an identity}
\label{main identity}

We recall that the function $\widehat{f}^{(N)}(\tau,z)$ is simply related to ${\cal N}=4$ massless character
\begin{eqnarray}
&&\hspace{-1cm}
 \widehat{\ch{(\stR)}{0}}(N-1, 0 ;\tau,z) \equiv (-1)^{N-1} \, 
\frac{2 \th_1(\tau,z)^2}{i\eta(\tau)^3\th_1(\tau,2z)}\, \hf^{(N)}(\tau,z).
\end{eqnarray}
Here,  
$\widehat{\ch{(\stR)}{0}}(N-1, 0 ;\tau,z)$ denotes the modular completion of the 
$\cN=4$ massless character of level $N-1$, isospin $0$ in the $\tR$-sector. 
This is actually the unique modular completion of  $\cN=4$ massless characters since they are  independent of the value of  isospin $\ell$,  
as was discussed in \cite{ES-nhP}.
We shall now prove the following important identity;
\begin{equation}
2 \frac{\th_1(\tau,z)}{\th_1(\tau,2z)} \, \hf^{(N)} (\tau,z)
= \sum_{\ell=0}^{N-2}\,
\sum_{a \in \bz_{N}}\, \ch{(\stR)}{\ell,\ell+1+2a}(\tau,z) \hF^{(N)}(\ell+1, a ; \tau, (N-1)z) .
\label{main id hf hF}
\end{equation}
This may be interpreted as a massless counterpart of \eqn{branching theta}, and 
it implies that
\eqn{ansatz cZ N=4 L}
is written in terms of the modular completion 
$\widehat{\ch{(\stR)}{0}}(N-1, 0 ;\tau,z)$.

~

\noindent
\underline{\bf Proof of \eqn{main id hf hF} : }

We set 
\begin{eqnarray}
&& G^{(\stR)}(\tau,z) := 2 \frac{\th_1(\tau,z)^2}{i \eta(\tau)^3 \th_1(\tau,2z)} \, \hf^{(N)} (\tau,z)
\nn
&& 
\hspace{2cm}
- \frac{\th_1(\tau,z)}{i \eta(\tau)^3}\, \sum_{\ell=0}^{N-2}\,
\sum_{a \in \bz_{N}}\, \ch{(\stR)}{\ell,\ell+1+2a}(\tau,z) 
\hF^{(N)}(\ell+1, a ; \tau, (N-1)z) ,
\label{def G tR}
\end{eqnarray}
and prove $G^{(\stR)} (\tau,z) \equiv 0$.

First of all, it is obvious by definition that $G^{(\stR)}(\tau,z)$
possesses the correct modular and spectral flow properties as a 
weak Jacobi form of weight 0, index $N-1$.

We next discuss more non-trivial properties of the function $G^{(\stR)}(\tau,z)$:

\newpage

\noindent
\underline{\bf (i) holomorphicity with respect to $\tau$ : }

By using \eqn{hF N}, \eqn{shadow R}
and 
\eqn{branching theta}, we obtain
\begin{eqnarray}
&& 
\hspace{-5mm}
\sqrt{\tau_2} \frac{\del}{\del \bar{\tau}} \left[
\sum_{\ell=0}^{N-2}\,
\sum_{a \in \bz_{N}}\, \ch{(\stR)}{\ell,\ell+1+2a}(\tau,z) 
\hF^{(N)}(\ell+1, a ; \tau, (N-1)z) 
\right]
\nn
&& 
\hspace{5mm}
= \frac{i}{4\sqrt{N}} \sum_{\ell=0}^{N-2} \sum_{a \in \bz_N} \sum_{j\in \bz_2} \,
\ch{(\stR)}{\ell,\ell+1+2a}(\tau,z) \, \overline{S_{\ell+1+Nj,N}(\tau)} \, \Th{\ell+1+Nj+2a}{N} \left(\tau, \frac{2(N-1)}{N}z\right)
\nn
&&
\hspace{5mm}
= \frac{i}{4 \sqrt{N}} \frac{2 \th_1(\tau,z)}{\th_1(\tau,2z)}\, \sum_{\ell=0}^{N-2} \,
\left[
\overline{S_{\ell+1,N}(\tau)} \, \Th{\ell+1}{N}^{[-]}(\tau,2z)
+ \overline{S_{N- \ell+1,N}(\tau)} \, \Th{N-\ell+1}{N}^{[-]}(\tau,2z)
\right]
\nn
&& 
\hspace{5mm}
= \frac{i}{4 \sqrt{N}} \frac{2 \th_1(\tau,z)}{\th_1(\tau,2z)}\, \sum_{r\in \bz_{2N}}\, \overline{S_{r,N}(\tau)} \, \Th{r}{N}(\tau,2z).
\label{shadow eval 1}
\end{eqnarray}
In the third line, we made use of the identities $ \ch{(\stR)}{\ell,m}(\tau,z) =  - \ch{(\stR)}{N-2-\ell,m+N}(\tau,z) $
and $S_{r+N,N}(\tau) = - S_{N-r,N}(\tau)$. 
Combining \eqn{shadow eval 1} with \eqn{shadow hf}, we conclude 
\begin{equation}
\frac{\del}{\del \bar{\tau}} G^{(\stR)}(\tau,z) =0.
\end{equation}

~


\noindent
\underline{\bf (ii) holomorphicity with respect to $z$}

We next confirm the holomorphicity with respect to $z$; in other words, the absence of singularities 
in $z$-variable. 
To this aim it would be useful to rewrite 
the function $G^{(\stR)}(\tau,z)$ in the form of non-holomorphic Poincare series \eqn{formula nhP 1};
\begin{eqnarray}
&& \hspace{-1.5cm}
G^{(\stR)}(\tau,z) = \frac{\th_1(z)^2}{2\pi \eta^3 \th_1(2z)} \,
\sum_{\la \in \La}\, \frac{\rho^{(N)}(\la,z)}{z+\la} 
\nn
&&  \hspace{-1.2cm}
- \frac{\th_1(z)}{2\pi \eta^3} \, \sum_{\la \equiv m\tau+n \in \La}\,
(-1)^{m+n} y^{\frac{N-2}{N}m} q^{\frac{N-2}{2N} m^2} e^{- 2\pi i \frac{mn}{N}}\, 
\frac{\th_1 \left(\frac{N-1}{N} (z+\la) \right)}{\th_1 \left(\frac{1}{N} (z+\la) \right)}
\, \frac{\rho^{(\frac{1}{N})}(-\la, (N-1) z)}{(N-1)z- \la},
\label{nh-P G}
\end{eqnarray}
where $\rho^{(\kappa)}(\la,z)$ has been defined by \eqn{def rho}.
The first term is obviously holomorphic. 

On the other hand, the potential singularities of the second term emerge  at 
the points
\begin{equation}
z = \frac{1}{N-1} \la, \hspace{1cm} (\any \la \in \La).
\label{pole}
\end{equation}
They  are, however,  canceled by  
the simple zeros coming from the factor $\th_1 \left(\frac{N-1}{N} (z+\la) \right)$
since
$$
\frac{N-1}{N} \left(\frac{\la}{N-1} +\la\right) =  \frac{N-1}{N} \frac{N \la}{N-1} = \la \in \La.
$$
Also, the factor $\th_1 \left(\frac{1}{N} (z+\la) \right)$
gives rise to  simple poles 
\begin{equation}
z= -\la + N \nu, \hspace{1cm} (\any \nu \in \La),
\end{equation} 
which are canceled by the remaining $\th_1(z)$.
Therefore we have confirmed the holomorphicity of $G^{(\stR)}(\tau,z)$.

~


\noindent
\underline{\bf (iii) IR behavior of $G^{(\sNS)}(\tau,z)$ around $\tau \sim i \infty$ :}

We next examine the IR-behavior of $G^{(\sNS)}(\tau,z)$, which is defined by the half spectral flow as
\begin{equation}
G^{(\sNS)}(\tau,z) := q^{\frac{N-1}{4}} y^{N-1} G^{(\stR)}\left(\tau, z + \frac{\tau+1}{2}\right).
\label{def G NS}
\end{equation}
The first term of \eqn{def G tR} is essentially equal to 
$\widehat{\ch{(\stR)}{0}}(N-1, \ell=0; \tau, z)$,
which is converted into 
$\widehat{\ch{(\sNS)}{0}}(N-1, \ell=N-1; \tau, z)$ by the half spectral flow, yielding the IR-behavior 
$
\sim q^{\frac{N-1}{4}}.
$

On the other hand, the 2nd term of \eqn{def G tR} yields (up to phases) 
$$
\sim \frac{\th_3(z)}{\eta^3} \,
\sum_{\ell=0}^{N-2} \, \sum_{a\in \bz_N} \, \ch{(\sNS)}{\ell, \ell+2a}(z)\,
q^{\frac{(N-1)^2}{4N}}\, y^{\frac{(N-1)^2}{N}}\, 
\hF^{(N)}\left(\ell+1, a ; \tau, (N-1)\left(z+\frac{\tau+1}{2} \right)\right).
$$
The leading contribution obviously comes from the term  of $\ell=0$, $a=0$, which gives
the IR-behavior
$$
\sim q^{-\frac{1}{8} - \frac{N-2}{8N} + \frac{(N-1)^2}{4N}+ \frac{N-1}{2N}} = q^{\frac{N-1}{4}}.
$$
In this way, we find 
\begin{equation}
G^{(\sNS)}(\tau,z) 
\sim \cO(q^{\frac{N-1}{4}}), \hspace{1cm} (\tau \sim i\infty).
\label{IR G NS}
\end{equation}

~


In summary, we have shown that $G^{(\stR)}(\tau,z)$ should be a holomorphic weak Jacobi form 
with weight 0, index $N-1$, and satisfies \eqn{IR G NS}.
This  is enough to conclude $G^{(\stR)}(\tau,z) \equiv 0$ because of the  lemma given in Appndix C.
\hspace{5mm} 
{\bf (Q.E.D)}

~


\subsubsection{Considerations on the effective central charge}

The above result for the elliptic genus of case 2 
$\cN=4$ Liouville theory is summarized as 
\begin{eqnarray}
\cZ^{\msc{case 2}} (\tau,z) 
&=& \frac{\th_1(\tau,z)}{i\eta(\tau)^3} \, \sum_{\ell=0}^{N-2}\,
\sum_{a \in \bz_{N}}\, \ch{(\stR)}{\ell,\ell+1+2a}(\tau,z) 
\hF^{(N)}(\ell+1, a ; \tau, (N-1)z)
\nn
& = & (-1)^{N-1} \, \widehat{\ch{(\stR)}{0}}(N-1, 0 ;\tau,z).
\label{formula cZ N=4 L}
\end{eqnarray}
Namely, we claim that {\em the elliptic genus of case 2 should be equal  
to the modular completion of the $\cN=4$ massless character 
$\ch{(\stR)}{0}(N-1, 0 ;\tau,z)$ itself.}
This has been suggested from the consideration on the holomorphic anomaly  
(or the shadow) given at subsection \ref{outline}. It is similar to the case of  
the elliptic genus of $\cN=2$ Liouville theory 
($\cong$ $SL(2)/U(1)$-supercoset) which is given only in terms of modular completions of 
$\cN=2$ massless characters \cite{Troost,ES-NH}. 


However, one might still ask: 
%
The holomorphic anomaly remains unchanged even if we add any holomorphic Jacobi form with weight 0, 
index $N-1$ to \eqn{formula cZ N=4 L}. How can we reject this possibility? 
In the simplest case of $N=2$, the absence of holomorphic Jacobi form ($\propto \phi_{0,1}(\tau,z)$)
just means the familiar fact of decoupling gravity in the non-compact models (see {\em e.g.} \cite{ES-BH}). 
However, for the cases with $N>2$, we have a number of holomorphic Jacobi forms, 
and the situation would get much more non-trivial.



To answer this question and confirm the validity of above result 
\eqn{formula cZ N=4 L}, let us  present a consideration about the effective central charge
mentioned before.
We start with refining the $c_{\msc{eff}}$-condition \eqn{cZ NS ceff}
based on the affine $SU(2)$-symmetry as the underlying structure of $\cN=4$ SCA, 
as we promised.
To exhibit it, we expand the elliptic genus in the NS-sector 
in terms of the angular variable $y\equiv e^{2\pi i z}$ as 
\begin{equation}
\cZ^{(\sNS)}(\tau,z) \equiv q^{\frac{N-1}{4}} y^{N-1} \, \cZ\left(\tau, z+ \frac{\tau+1}{2}\right)
= \frac{1}{2} \sum_{s \in \bz_{\geq 0}}  \, (y^s + y^{-s})  \cZ_s^{(\sNS)}(\tau) .
\label{cZ NS exp y}
\end{equation}


Then, we obtain the constraint;
\begin{equation}
\lim_{\tau_2 \,\rightarrow\, +\infty}\, e^{- 2\pi \tau_2 
\left\{ \frac{c_{\msc{eff}}}{24} - \frac{\ell \left(\ell+2 \right)}{4N}\right\}}
\, \left|\cZ_{s}^{(\sNS)}(\tau)\right| 
\equiv \lim_{\tau_2 \,\rightarrow\, +\infty}\, e^{- 2\pi \tau_2 
\left\{\frac{1}{4} - \frac{(\ell +1)^2}{4N} \right\}}
\, \left|\cZ_s^{(\sNS)}(\tau) \right|
< \infty,
\label{cZ NS ceff 2}
\end{equation}
where $\ell=0, 1, \ldots, N-1$ is defined by
\begin{equation}
\ell := 
\left\{
\begin{array}{ll}
|s|, & ~~ (0\leq |s| \leq N-2) \\
N-1, & ~~ (|s| \geq N-1)
\end{array}
\right.
\end{equation}


In fact, 
\begin{itemize}
\item $|s| = \ell \leq N-2$ :

The leading term contributing to $\cZ_s^{(\sNS)}(\tau)$ is 
composed of  the highest weight state of spin $\ell/2$ of bosonic $SU(2)$-current $j^a$, 
and the NS vacuum of fermions. Thus, the condition \eqn{cZ NS ceff 2} 
is obviously satisfied. 

\item $|s| \geq  N-1 $ :

The leading term contributing to $\cZ_s^{(\sNS)}(\tau)$ is 
composed of  the highest weight state of spin $\frac{N-2}{2}$ of $j^a$, 
and the level $\geq \frac{1}{2}$ state of  fermions. 
We thus obtain the IR-evaluation as 
$$
\cZ_s^{(\sNS)}(\tau) \sim q^{\al}, \hspace{1cm}
\al \geq \frac{(N-2)N}{4N} + \frac{1}{2} -  \frac{c_{\msc{eff}}}{24} \ge \frac{N^2-1}{4N} - \frac{c_{\msc{eff}}}{24}. 
$$
Therefore, 
the condition \eqn{cZ NS ceff 2} is still satisfied. 

\end{itemize}


Now, let us expand the holomorphic part of elliptic genus in the NS-sector in terms of $\cN=4$
characters;
\begin{eqnarray}
&& 
\cZ^{(\sNS)}_{\msc{hol}} (\tau,z) = \sum_{\ell=0}^{N-1} \, a_{\ell} \, \ch{(\sNS)}{0}(N-1, \ell ; \tau,z)
\nn
&& 
\hspace{2cm}
+ \sum_{j=0}^{N-2} \, \sum_{n \in \bz_{\geq 0}} \, b_{j,n} \, \ch{(\sNS)}{} \left(N-1, h= n + \frac{N-2}{4},  j ; \tau,z\right).
\label{ch exp cZ NS}
\end{eqnarray}
Here, the holomorphic part $\cZ^{(\sNS)}_{\msc{hol}} (\tau,z)$ has been defined 
so that it shows the same IR-behavior as $\cZ^{(\sNS)} (\tau,z)$ for each terms of $y$-expansions\footnote
   {In fact, the coefficients of massless characters $a_{\ell}$ can be uniquely determined 
from $\cZ^{(\sNS)} (\tau,z)$ by this assumption, 
while the massive coefficients $b_{j,n}$ would not be necessarily unique.
However, this ambiguity does not affect the following discussions. 
}.

It is easy to confirm that the constraint \eqn{cZ NS ceff 2} imposes that the conformal weight $h$ of 
the massive representations should satisfy the inequality
\begin{equation}
h \geq h(j) \equiv \frac{j(j+2)}{4N} + \frac{(N-1)^2}{4N} .
\label{ceff cond massive}
\end{equation}
It is notable that the second term $\frac{(N-1)^2}{4N}$ is equal the mass gap $\frac{\cQ^2}{8}$ due to the linear dilaton.

On the other hand, the IR-behavior of the NS massless character of isospin $\ell/2$ is evaluated as 
 $(\ell=0, \ldots, N-1)$
\begin{equation}
\ch{(\sNS)}{0}(N-1, \ell ; \tau,z)  \sim  q^{\frac{\ell}{2}-\frac{N-1}{4}},
\hspace{1cm} (\tau \sim i\infty).
\end{equation}
Thus, the constraint \eqn{cZ NS ceff 2} implies 
\begin{eqnarray}
&& \frac{\ell}{2}-\frac{N-1}{4} - \frac{(\ell+1)^2}{4N} + \frac{1}{4} \geq 0,
\nn
&&
\hspace{1cm}
\Longleftrightarrow ~ (\ell+1-N )^2 \leq 0.
\label{cZ NS ceff massless}
\end{eqnarray}
That is, only the maximal spin massless representation $\ell = N-1$ is allowed.

Consequently, \eqn{cZ NS ceff 2} implies that 
\begin{eqnarray}
&& 
\hspace{-5mm}
\cZ^{\msc{case 2} ~ (\sNS)}_{\msc{hol}} (\tau,z)  = 
a_{N-1} \,  \, \ch{(\sNS)}{0}(N-1, \ell=N-1 ; \tau,z)
\nn
&& 
\hspace{5mm}
+ \sum_{j=0}^{N-2} \, \sum_{n \in \bz_{\geq 0}, \,
h \geq h(j)
} \, b_{j,n} \, \ch{(\sNS)}{} 
\left(N-1, h= n+\frac{N-2}{4} , j ; \tau,z\right),
\label{ch exp cZ NS eval}
\end{eqnarray}
or equivalently, 
\begin{eqnarray}
&& 
\hspace{-5mm}
\cZ^{\msc{case 2} ~ (\stR)}_{\msc{hol}} (\tau,z)  = 
a'_0 \,  \, \ch{(\stR)}{0}(N-1, \ell=0 ; \tau,z)
\nn
&& 
\hspace{5mm}
+ \sum_{j=1}^{N-1} \, \sum_{n \in \bz_{\geq 0}, 
h\geq h(j-1) + \frac{1}{4}
} \, b'_{j,n} \, \ch{(\stR)}{} 
\left(N-1, h= n+ \frac{N-1}{4}, j ; \tau,z\right).
\label{ch exp cZ tR}
\end{eqnarray}
The above result \eqn{formula cZ N=4 L} is indeed consistent with this character expansion \eqn{ch exp cZ tR}
(with vanishing coefficients $b'_{j,n}$).

On the other hand, as shown in \cite{umbral1}, 
any holomorphic Jacobi form can never be written in the form \eqn{ch exp cZ tR}:
{\em i.e.}  we need additional contributions 
from the massless characters with $\ell \geq 1$, 
or the massive characters 
with conformal weight `below the massgap' $h< h(j-1)+ \frac{1}{4}$
in order to construct a holomorphic Jacobi form.   
In this way, we conclude that 
there is no room for adding extra holomorphic Jacobi forms to \eqn{formula cZ N=4 L}.

~


\subsubsection{Some remarks}

We add a few remarks for the analyses in this subsection:
\begin{description}
\item[(i)] 
It should be emphasized that  the case 2 is {\em not\/} equivalent 
with 
\begin{equation}
\left.
SU(2)_N/U(1) 
\otimes [\mbox{$\cN=2$ Liouville}]_{\cQ = -(N-1)\sqrt{\frac{2}{N}}}
\right|_{\bz_N-\msc{orbifold}},
\label{GKP}
\end{equation}
which is the superconformal system of the type studied in \cite{GKP}. 
The same value of linear dilaton $\cQ = -(N-1)\sqrt{\frac{2}{N}}$ 
is expected  for our case 2, but the $\cN=2$ Liouville potential does not preserve the $\cN=4$
superconformal symmetry except for the special case $N=2$ ($\hc =2$). 
In fact,
we have 
\begin{equation}
\hchid^{(N, (N-1)^2)}(v,a ;\tau,z) \equiv \frac{\th_1(\tau,z)}{i\eta(\tau)^3} \, \hF^{(N, (N-1)^2)}(v,a ;\tau,z),
\end{equation}
as the suitable building blocks for the $\cN=2$ Liouville theory with $\cQ = -(N-1)\sqrt{\frac{2}{N}}$,
which is equivalent to $SL(2)_{k=\frac{N}{(N-1)^2}}/U(1)$-supercoset.
The elliptic genus of \eqn{GKP} is obtained as 
\begin{eqnarray}
\hspace{-1cm}
\cZ(\tau,z)
& = & 
-\frac{\th_1(\tau,z)}{i\eta(\tau)^3} \,
\sum_{\ell=0}^{N-2} \, \sum_{a\in \bz_N} \, \ch{(\stR)}{\ell, \ell+1+2a} (\tau,z) \,
\hF^{(N, (N-1)^2)}(\ell+1,  a ;\tau, - z),
\label{cZ GKPelc} 
\end{eqnarray}
which is similar to \eqn{cZ ALE}.
On the other hand, \eqn{formula cZ N=4 L} can be rewritten as 
\begin{eqnarray}
&&
\cZ^{\msc{case 2}} (\tau,z) 
= 
-\frac{\th_1(\tau,z)}{i\eta(\tau)^3} \,
\sum_{\ell=0}^{N-2} \, \sum_{a\in \bz_N} \, \ch{(\stR)}{\ell, \ell+1+2a} (\tau,z) \,
\nn
&& \hspace{1cm}
\times \frac{1}{N-1} 
\sum_{\la \in \La/ (N-1) \La}\,
s^{(\frac{(N-1)^2}{N})}_{\la}
\cdot 
\hF^{(N, (N-1)^2)}(\ell+1,  a ;\tau, - z),
\label{formula cZ N=4 L 2} 
\end{eqnarray}
where $s^{(\kappa)}_{\la}$ denotes the spectral flow operator defined by \eqn{def sflow op}.
Stated physically, the $\bz_{N-1}$-orbifolding (or the Eichler-Zagier operator \cite{Eich-Zagier}) 
$$
\cW(N-1) \equiv \dsp \frac{1}{N-1} \sum_{\la \in \La/(N-1) \La} s^{(*)}_{\la} ,
$$ 
reduces the radius of asymptotic cylindrical region of the $\cN=2$ Liouville theory so as to be compatible 
with the $\cN=4$ superconformal symmetry.


\item[(ii)] 
An important consequence of  the `$c_{\msc{eff}}$-condition' \eqn{cZ NS ceff 2}, which physically means the normalizability of spectrum,
is the {\em inevitable\/} emergence of  holomorphic anomaly in $\cZ^{\msc{case 2}} (\tau,z) $
as long as we require its good modular property. 

It would be also worthwhile to note the fact that the  formulas of  modular S-transformations of 
any $\cN=4$ superconformal characters in the $\tR$-sector are schematically written as 
\begin{equation}
\left. [\mbox{$\cN=4$ character}]\right|_{S} = A\,  \ch{(\stR)}{0} (\ell=0) + 
\sum_j \int _{h \geq h(j)}\, [\mbox{$\cN=4$ massive character}],
\label{N=4 S-tr general}
\end{equation}
where the coefficient $A$ in the R.H.S 
does not vanish when the character of L.H.S is any massless character.
Namely, we find the following facts for the representations  in the R.H.S of \eqn{N=4 S-tr general}; 
\begin{itemize}
\item Only the $\ell=0$ massless character can appear.

\item The massive characters with conformal weight satisfying $h \geq h(j) $ (`above mass gap') 
only appears.

\end{itemize}
These representations are precisely identical to those satisfying the condition \eqn{cZ NS ceff 2}.
This feature is indeed anticipated, since the modular invariant partition function (and thus the elliptic genus) 
should only include  the normalizable spectrum that contributes to the net degrees of freedom.

\end{description}

~


\section{`Duality' in $\cN=4$ Liouville Theory and Umbral Moonshine}

In the previous sections 
we studied two systems possessing $\cN=4$ superconformal symmetry 

\begin{description}
\item[case 1: ($\hc=2$)] 

~

The elliptic genus is given as \eqn{cZ ALE};
$$
\cZ^{\msc{case 1}}(\tau,z) 
=  
-\frac{\th_1(\tau,z)}{i\eta(\tau)^3} \,
\sum_{r=1}^{N-1} \, \sum_{a\in \bz_N} \, \ch{(\stR)}{r-1, r+2a} (\tau,z) \,
\hF^{(N)}(r,  a ;\tau, - z),
$$

\item[case 2: ($\hat{c}=2(N-1)$)]

~

The elliptic genus is given as \eqn{formula cZ N=4 L};
\begin{eqnarray}
\cZ^{\msc{case 2}} (\tau,z) 
&=&  \frac{\th_1(\tau,z)}{i\eta(\tau)^3} \, \sum_{r=1}^{N-1}\,
\sum_{a \in \bz_{N}}\, \ch{(\stR)}{r-1,r+2a}(\tau,z) 
\hF^{(N)}(r, a ; \tau, (N-1)z)
\nn
& = & (-1)^{N-1} \, \widehat{\ch{(\stR)}{0}}(N-1, 0 ;\tau,z) .
\nonumber
\end{eqnarray}

\end{description}

As we emphasized several times, these two $\cN=4$ systems have the same degrees of freedom
even though the central charges differ from each other, and  are expected to be dual in the sense of 
$\mbox{AdS}_3/\mbox{CFT}_2$-correspondence. 
We can explicitly observe a very simple correspondence 
\begin{equation}
\hF^{(N)}(v,a ;\tau,-z) ~ \mbox{\bf for case 1} 
~ \longleftrightarrow ~ \hF^{(N)}(v,a ;\tau, (N-1)z) ~ \mbox{\bf for case 2}.
\label{correspondence}
\end{equation}


Now, we present some comments on the relation with the analyses on 
the `umbral moonshine' \cite{umbral1,umbral2,CH,HM,HMN}. 
In \cite{CH}, the authors studied (the holomorphic part of)
the extention of \eqn{cZ ALE} with general modular coefficients determined by 
the simply-laced root system $X$ corresponding to each Niemeier lattice. 
We have $\mbox{rank} \, X =24$ by definition, and let $N$ be the Coxetor number of $X$.  
A Niemeier lattice is explicitly expressed as  
\begin{equation}
X = \coprod_i \, X_i, \hspace{1cm} \sum_i \, \mbox{rank}\, X_i = 24,
\end{equation}
where each $X_i$ is the irreducible component of root system
possessing the common Coxetor number $N$.

We can schematically write 
\begin{eqnarray}
\cZ_X^{[\hc=2]} (\tau,z) & \equiv  & \sum_i \, \cZ^{\msc{case 1}(X_i)} (\tau,z)
\nn
& : = & 
-\frac{\th_1(\tau,z)}{i\eta(\tau)^3} \,
\sum_i\,
\sum_{r_i, s_i=1}^{N-1} \, \sum_{a_i \in \bz_N} \, \cN^{X_i}_{r_i, s_i } \, 
\ch{(\stR)}{r_i-1, s_i+2a_i} (\tau,z) \,
\hF^{(N)}(s_i,  a_i ;\tau, - z)
\nn
& = & 
-\frac{\th_1(\tau,z)}{i\eta(\tau)^3} \,
\sum_{r, s=1}^{N-1} \, \sum_{a \in \bz_N} \, \cN^{X}_{r, s} \, 
\ch{(\stR)}{r-1, s+2a} (\tau,z) \,
\hF^{(N)}(s,  a ;\tau, - z)
\label{cZ ALE X} 
\end{eqnarray}
where $\cN^{X_i}_{r,s}$ denotes the modular invariant coefficients of $SU(2)_{N-2}$ 
associated to the simply-laced root system $X_i$, and we set 
$\cN^X_{r,s} \equiv \sum_i\, \cN^{X_i}_{r,s}$.
One may identify 
$\cZ^{\msc{case 1}(X_i)} (\tau,z)$ as the elliptic genus 
of the ALE space  associated to the simple singularity of the type $X_i$. 
In \cite{CH} it was suggested that the root system 
$X= \coprod_i X_i$ should be identified as the geometrical data of 
various K3-singularities.

Since we assume $\mbox{rank} \, X \left( \equiv \sum_i \, \mbox{rank}\, X_i\right) =24$, 
we can rewrite \eqn{cZ ALE X} as 
\begin{equation}
\cZ_X^{[\hc=2]} (\tau,z) = 2 \phi_{0,1}(\tau,z) 
- \frac{\th_1(\tau,z)^2}{\eta(\tau)^3} \, \hh^X(\tau),
\label{cZ ALE X 2}
\end{equation}
where
$\hh^X(\tau)$ is the completion of mock modular form of weight $1/2$ characterized 
by the shadow 
\begin{eqnarray}
\sqrt{\tau_2} \frac{\del}{\del \bar{\tau}} \hh^X(\tau) 
& = & - \frac{i}{2 \sqrt{N}}  \sum_{r, s=1}^{N-1}\,   \cN^X_{r, s } \,
\chi^{(N-2)}_{r-1} (\tau,0)  \overline{S_{s,N}(\tau)}
\nn
&\equiv& - \frac{i}{2 \sqrt{N}} \frac{1}{\eta(\tau)^3} \sum_{r, s=1}^{N-1}\,  \cN^X_{r, s } \,
S_{r,N}(\tau) \overline{S_{s,N}(\tau)}.
\label{shadow hhX}
\end{eqnarray}
This 
is derived from \eqn{cZ ALE X} with the help of  the branching relation \eqn{branching ALE}.
(We recall that the space of weight 0, index 1 holomorphic Jacobi form is one-dimensional and spanned by $\phi_{0,1}$).


Let us next consider the type $X$ generalization of \eqn{formula cZ N=4 L},
which is related to \eqn{cZ ALE X} via the `duality correspondence' \eqn{correspondence}; 
\begin{eqnarray}
\hspace{-1cm}
&&
\cZ_X^{[\hc=2(N-1)]}(\tau,z) : =  
\frac{\th_1(\tau,z)}{i\eta(\tau)^3} \,
\sum_{r, s=1}^{N-1} \, \sum_{a \in \bz_N} \, \cN^X_{r, s} \, 
\ch{(\stR)}{r-1, s+2a} (\tau,z) \,
\hF^{(N)}(s,  a ;\tau, (N-1) z).
\label{cZ N=4 X} 
\end{eqnarray}


Another useful realization of the duality correspondence 
is given as the natural extension of the identity
\eqn{formula cZ ALE};
\begin{eqnarray}
\cZ_X^{[\hc=2]}(\tau,z) &=& 2 \phi_{0,1}(\tau,z) + 
\th_1(\tau,z)^2 \, \frac{1}{2\pi i} \oint_{w=0} \frac{dw}{w}\, 
\frac{e^{(N-2) G_2(\tau) w^2}}{\th_1(\tau,w)^2} \, \cZ_X^{[\hc=2(N-1)]}(\tau,w)
\nn
& \equiv & 2 \phi_{0,1}(\tau,z) 
+ \frac{\th_1(\tau,z)^2}{\eta(\tau)^6} \, \frac{1}{8\pi^3 i} \oint_{w=0}\frac{dw}{w}\, 
\frac{e^{(N-1)G_2(\tau) w^2}}{\sigma(\tau,w)^2} \, \cZ_X^{[\hc=2(N-1)]}(\tau,w),
\label{duality rel}
\end{eqnarray}
where we introduced the Weierstrass $\sigma$-function \eqn{w sigma} in the second line.
In fact, one can straightforwardly confirm that the second term possesses the correct modular property and reproduces 
the expected shadow \eqn{shadow hhX} in the  manner similar to the derivation of  \eqn{formula cZ ALE}.


Similarly to \eqn{cZ ALE X 2}, the R.H.S of  \eqn{cZ N=4 X}  can be decomposed as  
\begin{eqnarray}
\cZ_X^{[\hc=2(N-1)]}(\tau,z)  & =  & 
 \Phi_{0, N-1}^X(\tau,z) -  \frac{\th_1(\tau,z)^2}{\eta(\tau)^3} \, \sum_{r=1}^{N-1}\, \hh_r^X(\tau) 
\chi^{(N-2)}_{r-1}(\tau,2z)
\nn
& \equiv &  \Phi_{0, N-1}^X(\tau,z) - \frac{2 \th_1(\tau,z)^2}{i \eta(\tau)^3 \th_1(\tau,2z)} \,
\sum_{r=1}^{N-1}\, \hh_r^X(\tau) \Th{r}{N}^{[-]}(\tau,2z).
\label{cZ N=4 X 2}
\end{eqnarray}
In the above expression $ \Phi_{0, N-1}^X(\tau,z)$ is a weak Jacobi form of weight 0, index $N-1$, 
which is holomorphic with respect to $\tau$, 
but generically {\em meromophic with respect to $z$}.
$\chi^{(k)}_{\ell}(\tau,z)$ is the affine $SU(2)$ character of level $k$, isospin $\ell/2$,  and 
$\hh^X_r(\tau)$ are the completions of vector valued mock modular forms of weight 1/2, 
whose shadow is given as 
\begin{eqnarray}
\sqrt{\tau_2} \frac{\del}{\del \bar{\tau}} \hh_r^X(\tau) = - \frac{i}{2 \sqrt{N}} \sum_{s=1}^{N-1}\,   
\cN^X_{r, s } \,
\overline{S_{s,N}(\tau)}.
\label{shadow hhXr}
\end{eqnarray}
The formula \eqn{shadow hhXr}  again follows from the definition \eqn{cZ N=4 X 2} and the identity
\eqn{branching theta}.

We note a subtlety in the decomposition \eqn{cZ N=4 X 2}. 
$\hh_r^X(\tau)$ is not necessarily determined  only from the  shadow \eqn{shadow hhXr}. 
This is in contrast to $\hh^X(\tau)$ in \eqn{cZ ALE X 2} which is indeed determined uniquely.
For a sufficiently large $N$, there exist non-trivial holomorphic weak Jacobi forms of weight 1, index $N$, 
which we denote here, say, $\psi_{1,N}(\tau,z)$.
Then, we may add 
$
\frac{2 \th_1(\tau,z)^2}{i \eta(\tau)^3 \th_1(\tau,2z)} \, \psi^{[-]}_{1,N}(\tau,z)
$
to  the first term of \eqn{cZ N=4 X 2}, while subtracting the same function from the second term. 
We can thus modify $\hh^X_r(\tau)$ in \eqn{cZ N=4 X 2} with keeping the shadow \eqn{shadow hhXr} unchanged.
To avoid this ambiguity,
 we should impose the {\em `optimal growth condition'};
\begin{equation}
\lim_{\tau\, \rightarrow\, i\infty}\, q^{\frac{1}{4N}} \left| \hh^X_r(\tau) \right| < \infty,  \hspace{1cm} (\any r=1, \ldots, N-1),
\label{optimal growth cond}
\end{equation}
according to \cite{umbral2}.
In fact, the above $\psi^{[-]}_{1,N}(\tau,z)$ can be expanded by the theta functions 
as 
\begin{equation}
\psi^{[-]}_{1,N}(\tau,z) = \sum_{r=1}^{N-1}\, \al_r(\tau) \Th{r}{N}^{[-]}(\tau,2z),
\label{exp psi-}
\end{equation}
and the holomorphic coefficients $\al_r(\tau)$ cannot satisfy the condition 
 \eqn{optimal growth cond} due to the theorem {\bf 9.7} of \cite{DMZ}. 
Consequently, we can remove the ambiguity in the decomposition \eqn{cZ N=4 X 2},
and can determine $\hh^X_r(\tau)$ as well as $\Phi^X_{0,N-1}(\tau,z)$ uniquely.


Now, substituting the decompositions \eqn{cZ ALE X 2}, \eqn{cZ N=4 X 2} into the formula
\eqn{duality rel}, we find
\begin{equation}
\hh^X(\tau) = \sum_{r=1}^{N-1} \, \hh^X_r(\tau) \chi^{(N-2)}_{r-1}(\tau,0) 
\equiv \frac{1}{\eta(\tau)^3}  \sum_{r=1}^{N-1} \, \hh^X_r(\tau) S_{r,N}(\tau),
\label{rel hhX}
\end{equation}
since the contour integral
$
\dsp
\oint \frac{dw}{w} \, \frac{e^{(N-1) G_2(\tau) w^2}}{\sigma(\tau,w)^2}\, \Phi^X_{0,N-1}(\tau,w)
$
should be a holomorphic modular form of weight 2, and thus vanishes\footnote
  {We remark that the optimal growth condition is not necessary  for the purpose of 
 proving  the identity \eqn{rel hhX}.
In other words, the ambiguity of $\hh_r^X(\tau)$ mentioned above does not spoil this relation: 
$
\dsp 
\sum_{r=1}^{N-1}\, \al_r(\tau) \chi^{(N-2)}_{r-1}(\tau,0) =0
$
always holds for the coefficients $\al_r(\tau)$ appearing in \eqn{exp psi-}.}.
This is the duality relation between 
the expansion coefficients of massive representations of 
 $\cZ^{[\hc=2]}_X (\tau,z)$ and  $\cZ^{[\hc=2(N-1)]}_X (\tau,z)$. 
 In the case of Mathieu moonshine ($N=2,X=A^{24}_{1}$) one has the self-dual situation 
 $\widehat{h}^{A^{24}_{1}}(\tau)=\widehat{h}_{r=1}^{A^{24}_{1}}(\tau)$. In general 
holomorphic parts of $\hh_r^X(\tau)$ should reproduce mock modular form of umbral moonshine
on which the umbral group $G_X$ should act \cite{umbral1}.


Let us discuss the relation of the present analysis with a closely related consideration given in \cite{CH}.
For this purpose it is convenient to introduce the meromorphic Jacobi form $\Psi^X_{1,N}(\tau,z)$
with weight 1, index N defined by
\begin{equation}
\Psi^X_{1,N}(\tau,z) := \frac{i\eta(\tau)^3 \th_1(\tau,2z)}{2 \th_1(\tau,z)^2}\, \Phi^X_{0, N-1}(\tau,z)
\equiv \hf^{(1)}(\tau,z) \Phi^X_{0, N-1}(\tau,z) ,
\end{equation}
following  \cite{umbral1,umbral2}.
Then, \eqn{cZ N=4 X 2} can be rewritten as 
\begin{equation}
\Psi^X_{1,N}(\tau,z) = 
\hf^{(1)}(\tau,z) \cZ^{[\hc=2(N-1)]}_X (\tau,z) 
+  \sum_{r=1}^{N-1}\, \hh_r^X(\tau) \Th{r}{N}^{[-]}(\tau,2z),
\label{cZ N=4 X 3}
\end{equation}
where the first and second terms in R.H.S correspond to the {\em polar} and {\em finite} parts
in the terminology of \cite{umbral1,umbral2}.

If $X$ includes only the $A$-type components, 
our results  are manifestly consistent with those given in \cite{CH}.
Namely, both  the polar and finite parts in  \eqn{cZ N=4 X 3} 
(the `case 2 theory' with $\hc=2(N-1)$)
separately 
correspond to the first and second terms in \eqn{cZ ALE X 2}
(the `case 1 theory' with $\hc=2$) 
under the transformation given in \cite{CH};  
\begin{eqnarray}
\varphi_{1, N} (\tau,z) ~ 
&\mapsto& ~ \varphi'_{0, N^2}(\tau,z) := \frac{\th_1(\tau, (N-1)z) \th_1(\tau, Nz)}{i\eta(\tau)^3 \th_1(\tau,z)} 
\varphi_{1, N} (\tau,z)
\nn 
  &\mapsto& ~ \varphi''_{0,1}(\tau, z) 
:= \frac{1}{N} \sum_{\la \in \La/N\La}\, s^{(N^2)}_{\la} \cdot \varphi_{0,N^2}' (\tau, z/N) .
\label{CH map}  
\end{eqnarray}
Here, we denoted a (holomorphic or non-holomorphic) weak Jacobi form of weight $w$ and index $d$ by 
the symbol  `$\varphi_{w,d}(\tau,z)$', and the second line 
is identified as the Eichler-Zagier operator $\cW(N)$ ($\bz_N$-orbifolding).
In fact, the first term in R.H.S of \eqn{cZ N=4 X 3} just becomes 
\begin{equation}
 \hf^{(1)}(\tau,z) \cZ^{[\hc=2(N-1)]}_X (\tau,z) = \frac{24}{N-1} \hf^{(N)}(\tau,z),
\label{rel cZ X A}
\end{equation}
by substituting  the identity \eqn{main id hf hF} into \eqn{cZ N=4 X}, 
and the map \eqn{CH map} converts \eqn{cZ N=4 X 3}  into the definition of $\cZ^{[\hc=2]}_X(\tau,z)$ itself. 
(Recall the formula of elliptic genus of the $\cN=2$ minimal model \eqn{cZ min}.)
Furthermore, the correspondence of finite parts by \eqn{CH map} is  
found to be equivalent with the relation \eqn{rel hhX} due to the branching relation 
\eqn{branching ALE}.
The fact \eqn{rel cZ X A} also implies that $\Psi^X_{1,N}(\tau,z)$ here coincides with 
the `umbral Jacobi form' constructed in \cite{umbral2}.
It is also obvious that \eqn{rel cZ X A} is consistent with our duality relation \eqn{duality rel}
because of the formula \eqn{formula cZ ALE}.


In the cases when $X$ includes $D$ or $E$-type components, however, 
$\Psi^X_{1,N}(\tau,z)$ generally differs from the umbral Jacobi form.
Indeed, according to the explicit construction of the umbral Jacobi form given in \cite{umbral2}, 
it would contain the $n$-torsion points with $n | N$ created by the Eichler-Zagier operator 
written schematically as 
$$
\cW_X \equiv \sum_{n_i | N}\, c_i \cW(n_i).
$$
On the other hand, $\Psi^X_{1,N}(\tau,z)$ given above possesses 
the $n'$-torsion points with $n' | (N-1)$, which inherits from the functions
 $\hF^{(N)}(r,a; \tau, (N-1)z)$  in the expression \eqn{cZ N=4 X}\footnote
   {Such torsion points are canceled out in the cases when $X$ is made up only of the $A$-type components,
 as we illustrated in subsection \ref{main identity}. }.

Nevertheless, if we replace $\Psi^X_{1,N}(\tau,z)$ with the umbral Jacobi form of \cite{umbral2} 
in the `elliptic genus' $\cZ_X^{[\hc=2(N-1)]} (\tau,z)$, the duality relation 
\eqn{duality rel} is still satisfied, because the difference of these meromorphic Jacobi forms 
just yields a vanishing contribution to the contour integral.

How about the correspondence \eqn{CH map}?
Indeed, one can show that the claim of \cite{CH} still holds 
even when we replace the umbral Jacobi form with $\Psi^X_{1,N}(\tau,z)$.
Namely,
$\hf^{(1)}(\tau,z)\cZ_X^{[\hc= 2(N-1)]}(\tau,z)$ 
and $\Psi^X_{1,N}(\tau,z)$ still correspond to 
the $\cZ^{[\hc=2]}_X(\tau,z)$ and $\cZ^{K3}(\tau,z) \equiv 2 \phi_{0,1}(\tau,z)$ by 
the map \eqn{CH map}, 
as shown by the following arguments;
\begin{description}
\item[(i)] 
It turns out that the finite part of \eqn{cZ N=4 X 3} is mapped  to that of $\cZ^{[\hc=2]}_X(\tau,z)$, 
 and the relation \eqn{rel hhX} still holds.

\item[(ii)] 
When considering the map \eqn{CH map}, 
all the $n'$-torsion points included in $\Psi^X_{1,N}(\tau,z)$ are canceled out 
with the zeros of the factor $\th_1(\tau, (N-1)z)$ \footnote
   {On the other hand, the $n$-torsion points appearing in 
the umbral Jacobi form are canceled out by the factor $\th_1(\tau, Nz)$.}. 
We thus find that $\Psi^X_{1,N}(\tau,z)$ is always mapped to the unique holomorphic Jacobi
form $\al \phi_{0,1}(\tau,z)$ with some constant coefficient $\al$.

\item[(iii)]
This coefficient $\al$ is, however, just determined from the finite part
so that the contribution of `graviton representation' should be canceled out.
In fact, it is straightforward to confirm that \eqn{cZ N=4 X} cannot include such a term
after converting it into the $\hc=2$-system by the transformation \eqn{CH map}. 
This is quite similar to the argument given in the subsection \ref{remark cZ ALE}
(see the discussions around \eqn{formula cZ NS IR eval 1}, \eqn{formula cZ NS IR eval 2}).
On the other hand, the finite part of \eqn{cZ N=4 X} is mapped to 
$\hh^X(\tau) \frac{\th_1(\tau,z)^2}{\eta(\tau)^3}$ as noted above,
yielding the $h=0$ behavior $\sim - \frac{N-1}{12} q^{-\frac{1}{4}}$ in the NS-sector
as in \eqn{formula cZ NS IR eval 2}. 
Therefore, the cancellation of the graviton term implies that 
$$
\al = \frac{24}{N-1} \cdot \frac{N-1}{12} = 2,
$$
which completes the proof of our statement.

\end{description}

~


\section{Discussions}

We conclude that our ansatz \eqn{cZ N=4 X} for $\cZ^{[\hc=2(N-1)]}_X(\tau,z)$  should reproduce the expected  expansion coefficients $\hh^X_r(\tau)$ whose holomorphic parts give the coefficients of massive representations of Mathieu and umbral moonshine.
Therefore, we naturally consider that 
there are two world-sheet descriptions of the moonshine phenomena 
based on the theories $\cZ^{[\hc=2]}_X(\tau,z)$ and $\cZ^{[\hc=2(N-1)]}_X(\tau,z)$, depending on a different choice of $SU(2)_R$ symmetry. 
Mathieu moonshine sits at the self-dual point.

Our observation is related to the work \cite{CH} 
(also to the paper \cite{HMN}). 
Without invoking the  idea of duality authors of \cite{CH}   
introduced the transformation rule  between corresponding Jacobi forms of the two theories which seem to fit to our results very well. 
Possible geometrical interpretation of the umbral moonshine  
based on particular types of singular  K3 is also discussed.
As we briefly discussed in section 4, 
the correspondence \eqn{CH map} proposed in \cite{CH} is likely to be consistent with our `duality relation' \eqn{duality rel}.


Some of the  results in this paper appear correct by symmetry arguments (modularity etc) but are not verified explicitly: 
We have not computed the holomorphic parts of functions 
 $\widehat{h}^X_r$ except the case of Mathieu moonshine. 
In subsequent work we want to fill these gaps. We also discussed mainly A-type modular invariant and not D and E types.

One should keep it in mind that the Jacobi form $\Phi^X_{0,N-1}(\tau,z)$ given here
is at most meromorphic except for the purely $A$-type models, since $\Psi^X_{1,N}(\tau,z)$ generally includes torsion points as we mentioned.
Consequently, it would be subtle whether one can strictly interpret 
$\cZ^{[\hc=2(N-1)]}_X(\tau,z)$ as the elliptic genus of 
a well-defined superconformal system. 
Of course, we have no such subtlety for the `dual' $\hc=2$-realization \eqn{cZ ALE X} 
for an arbitrary $X$.  
We would like to further discuss this point in a future work.


\vskip2cm

\noindent
{\bf Acknowledgments}

We thank the organizers of LMS-EPSRC Durham Symposium on `New Moonshines, Mock Modular Forms and String Theory',
at Durham Univ. for August 3-12, 2015 for kind hospitality, where part of this work was done. 

Research of T.E. is supported in part by JSPS KAKENHI grant no. 25400273,
22224001 and 23340115.
Research of Y.S. is supported in part by JSPS KAKENHI grant no. 23540322.


\newpage

\appendix

\section*{Appendix A: ~ Notations and Useful Formulas}

\setcounter{equation}{0}
\def\theequation{A.\arabic{equation}}


In Appendix A we summarize the notations adopted in this paper and related useful formulas.
We assume $\tau\equiv \tau_1+i\tau_2$, $\tau_2>0$ and 
 set $q:= e^{2\pi i \tau}$, $y:=e^{2\pi i z}$;


~

\begin{description}

\item[\underline{Theta functions :}]
%
%
 \begin{equation}
 \begin{array}{l}
 \dsp \th_1(\tau,z)=i\sum_{n=-\infty}^{\infty}(-1)^n q^{(n-1/2)^2/2} y^{n-1/2}
  \equiv 2 \sin(\pi z)q^{1/8}\prod_{m=1}^{\infty}
    (1-q^m)(1-yq^m)(1-y^{-1}q^m), \\
 \dsp \th_2(\tau,z)=\sum_{n=-\infty}^{\infty} q^{(n-1/2)^2/2} y^{n-1/2}
  \equiv 2 \cos(\pi z)q^{1/8}\prod_{m=1}^{\infty}
    (1-q^m)(1+yq^m)(1+y^{-1}q^m), \\
 \dsp \th_3(\tau,z)=\sum_{n=-\infty}^{\infty} q^{n^2/2} y^{n}
  \equiv \prod_{m=1}^{\infty}
    (1-q^m)(1+yq^{m-1/2})(1+y^{-1}q^{m-1/2}), \\
 \dsp \th_4(\tau,z)=\sum_{n=-\infty}^{\infty}(-1)^n q^{n^2/2} y^{n}
  \equiv \prod_{m=1}^{\infty}
    (1-q^m)(1-yq^{m-1/2})(1-y^{-1}q^{m-1/2}) .
 \end{array}
\label{th}
 \end{equation}
\begin{eqnarray}
 \Th{m}{k}(\tau,z)&=&\sum_{n=-\infty}^{\infty}
 q^{k(n+\frac{m}{2k})^2}y^{k(n+\frac{m}{2k})} ,
\\
\Theta^{[-]}_{m,k} (\tau,z) & = & \frac{1}{2} \left[\Th{m}{k}(\tau,z) - \Th{m}{k} (\tau,-z) \right] .
 \end{eqnarray}
 We use abbreviations; $\th_i (\tau) \equiv \th_i(\tau, 0)$
 ($\th_1(\tau)\equiv 0$), 
$\Th{m}{k}(\tau) \equiv \Th{m}{k}(\tau,0)$.
 We also set
 \begin{equation}
 \eta(\tau)=q^{1/24}\prod_{n=1}^{\infty}(1-q^n).
 \end{equation}
%
%
The spectral flow properties of theta functions are summarized 
as follows;
\begin{eqnarray}
 && \th_1(\tau, z+m\tau+n) = (-1)^{m+n} 
q^{-\frac{m^2}{2}} y^{-m} \th_1(\tau,z) , \nn
&& \th_2(\tau, z+m\tau+n) = (-1)^{n} 
q^{-\frac{m^2}{2}} y^{-m} \th_2(\tau,z) , \nn
&& \th_3(\tau, z+m\tau+n) = 
q^{-\frac{m^2}{2}} y^{-m} \th_3(\tau,z) , \nn
&& \th_4(\tau, z+m\tau+n) = (-1)^{m} 
q^{-\frac{m^2}{2}} y^{-m} \th_4(\tau,z) , 
\nn
&& \Th{a}{k}(\tau, 2(z+m\tau+n)) = 
e^{2\pi i n a} 
q^{-k m^2} y^{-2 k m} \Th{a+2km}{k}(\tau,2z)~.
\label{sflow theta}
\end{eqnarray}

We introduce the Weierstrass $\sigma$-function;
\begin{eqnarray}
\sigma(\tau,z) & := & 
e^{\frac{1}{2} G_2(\tau) z^2}\,
\frac{\th_1(\tau,z)}{2\pi \eta(\tau)^3}
\nn
& \equiv & z \prod_{\om \in \La'} \, \left(1-\frac{z}{\om} \right)\, 
e^{\frac{z}{\om} + \frac{1}{2} \left(\frac{z}{\om}\right)^2}.
\hspace{1cm} \left(\La' \equiv \La - \{ 0\}\right),
\label{w sigma}
\end{eqnarray}
where $G_2(\tau)$ is the (unnormalized) second Eisenstein series;
\begin{eqnarray}
G_2(\tau) & := & \sum_{n \in \bz-\{0\}}\, \frac{1}{n^2} + \sum_{m\in \bz-\{0\}}\, \sum_{n\in \bz} \, \frac{1}{(m\tau+n)^2}
\nn
& \equiv & \frac{\pi^2}{3} 
\left[1 - 24 \sum_{n=1}^{\infty}\, \frac{n q^n}{1-q^n} 
\right],
\label{def G2}
\end{eqnarray}
It is useful to note the anomalous $S$-transformation formula of $G_2(\tau)$;
\begin{equation}
G_2 \left(- \frac{1}{\tau} \right) = \tau^2 G_2(\tau) -2\pi i \tau.
\label{G2 S}
\end{equation}
We also set
\begin{equation}
\hG_2(\tau) := G_2(\tau) -\frac{\pi}{\tau_2},
\label{def hG2}
\end{equation}
which is a non-holomorphic modular form of weight 2.

~

\item[\underline{Spectral flow operator :}]
(see also \cite{Eich-Zagier})
\begin{eqnarray}
s^{(\kappa)}_{\la} \cdot f(\tau,z) &:= & e^{2\pi i \frac{\kappa}{\tau_2} \la_2 \left(\la + 2z\right) }\,
f(\tau, z+\la)
\nn
& \equiv &q^{\kappa \al^2} y^{2\kappa \al} e^{2\pi i \kappa \al \beta}\,
f(\tau,z+\al \tau+ \beta) ,
\nn
&& 
\hspace{4cm} 
(\la \equiv \al \tau +\beta, ~ \any \al, \beta \in \br).
\label{def sflow op}
\end{eqnarray}

An important property of the spectral flow operator $s^{(\kappa)}_{\la}$ is 
the modular covariance, which precisely means the following:
\\
Assume that $f(\tau,z)$ is an arbitrary function with the modular property;
$$
f(\tau+1,z) = f(\tau,z), \hspace{1cm} 
f\left(-\frac{1}{\tau}, \frac{z}{\tau}\right) = e^{2\pi i \frac{\kappa}{\tau} z^2} \tau^{\al} \, f(\tau,z),
$$
then, we obtain for $\any \la \in \bc$
$$
s^{(\kappa)}_{\la} \cdot f(\tau+1,z) = 
s^{(\kappa)}_{\la} \cdot f(\tau,z), 
\hspace{1cm} 
s^{(\kappa)}_{\frac{\la}{\tau}} \cdot f\left(-\frac{1}{\tau}, \frac{z}{\tau}\right) 
= e^{2\pi i \frac{\kappa}{\tau} z^2} \tau^{\al}\, s^{(\kappa)}_{\la} \cdot f(\tau,z).
$$ 

~


\item[\underline{Error functions} :]
\begin{eqnarray}
\erf(x) &:=& \frac{2}{\sqrt{\pi}} \int_0^{x} e^{-t^2} dt,  \hspace{1cm} (x\in \br),
\label{Erf}
\\
\erfc(x) &:=&   \frac{2}{\sqrt{\pi}} \int_{x}^{\infty} e^{-t^2} dt
\equiv 1 - \erf(x) ,
\hspace{1cm} (x>0)
\label{Erfc}
\end{eqnarray}
%
The next identity is elementary but  useful;
\begin{eqnarray}
&& 
\hspace{-2cm}
\sgn(\nu +0) -  \erf(\nu ) 
= \sgn(\nu+0) \erfc (|\nu| ) 
= \frac{1}{i\pi} 
\int_{\br - i0}\, dp \, \frac{e^{-(p^2 + \nu^2)}}{p-i\nu}.
\hspace{1cm} (\nu \in \br),
\label{id erf}
\end{eqnarray}


~

\item[\underline{weak Jacobi forms} :]

~

The weak Jacobi form \cite{Eich-Zagier} for 
the full modular group $\Gamma(1) \equiv SL(2,\bz)$
with weight $k (\in \bz_{\geq 0})$ and index
$r (\in \frac{1}{2} \bz_{\geq 0})$ 
is defined by the conditions 
\begin{description}
 \item[(i) modularity :] 
\begin{eqnarray}
 && \hspace{-1cm}
\Phi\left(
\frac{a\tau+b}{c\tau+d}, \frac{z}{c\tau+d}
\right) 
= e^{2\pi i r \frac{c z^2}{c\tau+d}} (c\tau+d)^k \, \Phi(\tau,z), ~~~
\any \left(
\begin{array}{cc}
 a& b\\
 c& d
\end{array}
\right) \in \Gamma(1).
\label{modularity J}
\end{eqnarray}
\item[(ii) double quasi-periodicity :] 
\begin{eqnarray}
 \Phi(\tau,z+m\tau+n) = (-1)^{2r(m+n)} q^{-r m^2} y^{-2r m}\, \Phi(\tau,z). 
\label{sflow J}
\end{eqnarray}
%
\end{description}
In this paper, we shall use this terminology in a broader sense. 
We allow a half integral index $r$, and more crucially,  allow non-holomorphic dependence on $\tau$,
while we keep the holomorphicity with respect to $z$ \footnote
{According to  the original terminology of \cite{Eich-Zagier},
the `weak Jacobi form' of weight $k$ and index $r$ ($k, r \in \bz_{\geq 0}$) means that $\Phi(\tau,z)$ should 
be Fourier expanded as 
$$
\Phi(\tau,z) = \sum_{n\in \bz_{\geq 0}}\, \sum_{\ell \in \bz}\, c(n,\ell) q^n y^{\ell},
$$
in addition to the conditions \eqn{modularity J} and \eqn{sflow J}.
It is called the `Jacobi form' if it further satisfies  the condition: 
$c(n,\ell) =0$ for $\any n, \ell$ s.t.  $4n r -\ell^2 <0$.
}.

\end{description}

~


\section*{Appendix B: ~ Derivation of the Formulas \eqn{formula hHN}, \eqn{formula HN 1} and \eqn{formula hH2} }
\label{app hH}

\setcounter{equation}{0}
\def\theequation{B.\arabic{equation}}


In this appendix, we derive the  formulas  \eqn{formula hHN},  \eqn{formula HN 1} and \eqn{formula hH2}.



\newpage

\noindent
\underline{\bf Derivation of \eqn{formula hHN} : }

We start with the definition of $\hH^{(N)}(\tau)$ \eqn{def hHN}. 
Substituting   \eqn{formula nhP 1} into \eqn{def hHN}, 
we obtain 
($\La \equiv \bz \tau +\bz$)
\begin{eqnarray}
&& \hH^{(N)}(\tau) 
 =  \frac{1}{2 \pi^2 } \oint_{w=0} \frac{dw}{w^2} \, 
\left[ \frac{ w \eta(\tau)^3}{i \th_1(\tau,2w)} e^{(N-2) G_2(\tau) w^2}
\, \sum_{\la \in \La} \, 
\frac{e^{-\frac{\pi}{\tau_2} N \left\{\left|\la \right|^2 + 2\bar{\la} w + w^2 \right\}}}{\la+w}
\right].
\label{eval hHN 0}
\end{eqnarray}
It would be easiest to carry out this residue integral 
by introducing  the Weierstrass $\sigma$-function  defined as \eqn{w sigma}.
Namely,
\begin{eqnarray}
\hspace{-5mm}
\hH^{(N)}(\tau) 
& = & \frac{1}{8 \pi^3 i} \oint_{w=0} \frac{dw}{w^2} \, 
\left[ \frac{ 2 w}{\sigma(\tau, 2w)} e^{N G_2(\tau) w^2}
\, \sum_{\la \in \La} \, 
\frac{e^{-\frac{\pi}{\tau_2} N \left\{\left|\la \right|^2 + 2\bar{\la} w + w^2 \right\}}}{\la+w}
\right]
\nn
& = & \frac{1}{8 \pi^3 i} \oint_{w=0} \frac{dw}{w^2} \, 
\left[ \frac{ 2 w}{\sigma(\tau, 2w)} e^{N G_2(\tau) w^2}
\, \sum_{\la \in \La'} \, 
\frac{e^{-\frac{\pi}{\tau_2} N \left\{\left|\la \right|^2 + 2\bar{\la} w + w^2 \right\}}}{\la+w}
\right]
\nn
&& \hspace{1cm} + \frac{1}{8 \pi^3 i} \oint_{w=0} \frac{dw}{w^3} \,
\frac{ 2 w}{\sigma(\tau, 2w)} e^{N \widehat{G}_2(\tau) w^2}.
\hspace{1cm}
\left(\La' \equiv \La -\{0\}\right).
\label{eval hHN 3}
\end{eqnarray}
Moreover, since we have
$$
\frac{2w}{\sigma(\tau,2w)} = 1 + \cO(w^4),
$$
due to \eqn{w sigma}, 
\eqn{eval hHN 3} just becomes
\begin{eqnarray}
\hH^{(N)}(\tau) & = & 
 \frac{1}{8 \pi^3 i} \oint_{w=0} \frac{dw}{w^2} \, 
\sum_{\la \in \La'} \, 
\frac{e^{-\frac{\pi}{\tau_2} N \left\{\left|\la \right|^2 + 2\bar{\la} w + w^2 \right\}}}{\la+w}
+ \frac{1}{8 \pi^3 i} \oint_{w=0} \frac{dw}{w^3} \,  e^{N \widehat{G}_2(\tau) w^2}
\nn
& = & \frac{1}{4\pi^2} \left[ \frac{\del}{\del w} \sum_{\la \in \La'}\, 
\left.
\frac{e^{-\frac{\pi}{\tau_2} N \left\{\left|\la \right|^2 + 2\bar{\la} w + w^2 \right\}}}{\la+w}
\right|_{w=0} + N \widehat{G}_2(\tau) \right]
\nn
&=& \frac{1}{4\pi^2} \left[  -  
\sum_{\la \in \La'} \, \frac{e^{- \frac{\pi}{\tau_2} N \left|\la\right|^2 }}{\la^2} 
\left\{
1+ \frac{2 \pi N}{\tau_2} \left|\la\right|^2
\right\} + N \hG_2(\tau)\right].
\label{eval hHN 4}
\end{eqnarray}
We have thus obtained \eqn{formula hHN}.

~


\noindent
\underline{\bf Derivation of \eqn{formula HN 1} : }

We next evaluate the holomorphic part of \eqn{formula hHN}
or \eqn{eval hHN 4}.
Let us derive the expression of the holomorphic function $H^{(N)}(\tau)$
given in \eqn{formula HN 1}.

Recalling \eqn{eval shadow ALE}, the non-holomorphic part of 
$\cZ^{\msc{case 1}}(\tau,z)$ can be rewritten as 
\begin{eqnarray}
\vD \cZ^{\msc{case 1}}(\tau,z) & \equiv & \frac{\th_1(\tau,z)^2}{\eta(\tau)^6} \, \vD \hH^{(N)}(\tau),
\nn
\vD \hH^{(N)}(\tau) & =& 2 \eta(\tau)^3 \, \lim_{w\rightarrow 0}\,
\left[\frac{\hf^{(N)} (w) - f^{(N)\,[-]}}{i\th_1(2w)}\right]
\nn
& = & \frac{1}{2\pi i}\, \frac{\del}{\del w} \left. 
\left[\hf^{(N)} (w) - f^{(N)\,[-]}\right]\right|_{w=0}.
\label{eval DhHN 1}
\end{eqnarray}
Moreover, we note the formula
\begin{equation}
\hf^{(N)} (w) - f^{(N)\,[-]}(w) = \frac{i}{2\pi} \sum_{\la \in \La}\, s^{(N)}_{\la} \cdot
\left[\frac{e^{-\frac{\pi}{\tau_2}N w^2}-1}{w}\right],
\label{nhP non-hol part}
\end{equation}
where $s^{(N)}_{\la}$ denotes the spectral flow operator \eqn{def sflow op}.

Substituting \eqn{nhP non-hol part} into \eqn{eval DhHN 1}, we obtain
\begin{eqnarray}
\vD \hH^{(N)}(\tau) & = & \frac{1}{4\pi^2} \, \lim_{ \ep \rightarrow + 0}\,
\left[ \frac{\del}{\del w} \left. \sum_{\la \in \La'}\, s^{(N)}_{\la} \cdot
\left\{ \frac{e^{-\frac{\pi}{\tau_2}N w^2}- e^{-\frac{\pi}{\tau_2}\ep w^2}}{w}\right\} \right|_{w=0} 
+ \frac{\del}{\del w} \left. \left\{ \frac{e^{-\frac{\pi}{\tau_2}N w^2}-1}{w}\right\} \right|_{w=0}
\right]
\nn
& = & \frac{1}{4\pi^2} \, 
\left[ \frac{\del}{\del w} \left. \left\{  \sum_{\la \in \La'}\, s^{(N)}_{\la} \cdot
\frac{e^{-\frac{\pi}{\tau_2}N w^2}}{w}\right\} \right|_{w=0} 
- \frac{\pi N}{\tau_2} \right]
\nn
&& \hspace{4cm} 
- \lim_{\ep \rightarrow + 0}\, \frac{1}{4\pi^2}\, 
\frac{\del}{\del w}  \left.
\left\{ \sum_{\la\in \La'}\, s_{\la}^{(N)} \cdot \frac{e^{-\frac{\pi}{\tau_2}\ep w^2}}{w} \right\} \right|_{w=0}
\nn
& = & 
\frac{1}{4\pi^2} \, 
\left[ \frac{\del}{\del w} \left. \left\{ \sum_{\la \in \La'}\, s^{(N)}_{\la} \cdot
 \frac{e^{-\frac{\pi}{\tau_2}N w^2}}{w}\right\} \right|_{w=0} 
- \frac{\pi N}{\tau_2} \right]
\nn
&& \hspace{2cm}
- \frac{1}{4\pi^2} \, \frac{\del}{\del w} \left. \left[   
\sum_{m\neq 0}\, {\sum_{n\in\bz}}^P\, s^{(N)}_{m\tau+n} \cdot \frac{1}{w} 
+ {\sum_{n\neq 0}}^P\, \frac{1}{w+n}
\right]\right|_{w=0} .
\label{eval vhHN 2}
\end{eqnarray}
Here, we included  the convergence factor $e^{-\frac{\pi}{\tau_2}\ep w^2}$  
to make the second line well-defined, and 
the symbol '${\sum_{n}}^P$' denotes the principal value defined in 
\eqn{def sumP};
$$
{\sum_{n\neq 0}}^P \, a_n := \lim_{N\,\rightarrow\, \infty} \sum_{n=1}^N \, \left(a_n + a_{-n}\right), 
\hspace{1cm} 
{\sum_{n}}^P \, a_n := a_0 + {\sum_{n\neq 0}}^P \, a_n.
$$
Then, 
by comparing \eqn{eval vhHN 2} with \eqn{eval hHN 4}, and 
recalling 
$
\dsp 
\hG_2 (\tau) = G_2(\tau) - \frac{\pi}{\tau_2},
$ 
we find that the holomorphic part 
$
H^{(N)}(\tau) \equiv \hH^{(N)}(\tau) - \vD \hH^{(N)}(\tau)
$
is indeed given by the formula \eqn{formula HN 1}.


~ 

\noindent
\underline{\bf Derivation of \eqn{formula hH2} : }

Finally, let us derive the formula given in the second line of \eqn{formula hH2} for the $N=2$ case. 
To this end, it is again convenient to make use of the $\sigma$-function \eqn{w sigma}.
Substituting the identity \eqn{formula nhP 2} into the first line of \eqn{formula hH2}, we obtain  
\begin{eqnarray}
\hH^{(2)}(\tau) & \equiv &  \frac{1}{2  \pi i } \oint_{w=0} \frac{dw}{w} \, 
\frac{ \eta(\tau)^3}{i \th_1(\tau,w)} \htf^{(1/2)}(\tau,w)
\nn
& = &  \frac{1}{8  \pi^3 i } \oint_{w=0} \frac{dw}{w^2} \,
\frac{ w }{\sigma(\tau,w )}\, e^{\frac{1}{2} G_2(\tau) w^2}\,
\sum_{\la \in \La} \, (-1)^{m+n+mn}\,
\frac{e^{-\frac{\pi}{2 \tau_2} \left\{\left|\la \right|^2 + 2\bar{\la} w + w^2 \right\}}}{\la+w}
\nn
& = & 
 \frac{1}{8 \pi^3 i} \oint_{w=0} \frac{dw}{w^2} \, e^{\frac{1}{2} G_2(\tau) w^2}
\, \sum_{\la = m\tau+n \in \La'} \, (-1)^{m+n+mn} 
\frac{e^{-\frac{\pi}{2 \tau_2}  \left\{\left|\la \right|^2 + 2\bar{\la} w + w^2 \right\}}}{\la+w}
\nn
&& \hspace{3cm} + \frac{1}{8 \pi^3 i} \oint_{w=0} \frac{dw}{w^3} \, e^{\frac{1}{2} \widehat{G}_2(\tau) w^2}
\nn
& = & \frac{1}{4\pi^2} 
\left. \left[ \frac{\del}{\del w} \,
\sum_{\la \in \La'} \, (-1)^{m+n+mn}\,
\frac{e^{-\frac{\pi}{2 \tau_2} \left\{\left|\la \right|^2 + 2\bar{\la} w + w^2 \right\}}}{\la+w}
\right|_{w=0} + \frac{1}{2} \hG_2(\tau) \right]
\nn
& = & \frac{1}{4\pi^2}\left[ 
\sum_{\la \in \La'}\,
(-1)^{m+n+mn}\,e^{- \frac{\pi}{2\tau_2} \left|\la\right|^2 } 
\left(
- \frac{1 }{\la^2} 
+ \frac{2\pi i m  }{\la} - \frac{\pi}{\tau_2}
\right)
+ \frac{1}{2} \hG_2(\tau) \right].
\nn
& = & \frac{1}{4\pi^2}\left[ 
\sum_{\la \in \La'}\,
(-1)^{m+n+mn}\,e^{- \frac{\pi}{2\tau_2} \left|\la\right|^2 } 
\left(
- \frac{1 }{\la^2} 
+ \frac{2\pi i m  }{\la}
\right)
+ \frac{\pi}{\tau_2}
+ \frac{1}{2} \hG_2(\tau) \right].
\label{eval hH2 0}
\end{eqnarray}
In the last line, we made use of the identity 
\begin{equation}
\sum_{\la =m\tau+n \in \La} \, (-1)^{m+n+mn}\, e^{-\frac{\pi}{2 \tau_2}\left|\la\right|^2} =0.
\label{identity 1}
\end{equation}
In fact, 
\begin{eqnarray}
\sum_{\la \in \La} \, (-1)^{m+n+mn}\, e^{-\frac{\pi}{2 \tau_2}\left|\la\right|^2} & = & 
\left( \sum_{m, n \in 2\bz} - \sum_{m \in 2\bz+1 \, \msc{or} \, n \in 2\bz+1} \right) \,  e^{-\frac{\pi}{2 \tau_2}\left|\la\right|^2}
\nn
& = & \sum_{\la \in \La} \,  e^{-\frac{2 \pi}{\tau_2}\left|\la\right|^2}
- \left\{ 
\sum_{\la \in \La} \,  e^{-\frac{\pi}{2 \tau_2}\left|\la\right|^2}
-\sum_{\la \in \La} \,  e^{-\frac{2 \pi}{\tau_2}\left|\la\right|^2}
\right\} 
\nn
& = & 2 \sum_{\la \in \La} \,  e^{-\frac{2 \pi}{\tau_2}\left|\la\right|^2} 
- \sum_{\la \in \La} \,  e^{-\frac{\pi}{2 \tau_2}\left|\la\right|^2},
\nonumber
\end{eqnarray}
and the last line vanishes due to the Poisson resummation  with respect to both of $m$, $n$.

To proceed further, we set 
\begin{equation}
g(\tau) : = - \sum_{\la \in \La'} \, (-1)^{m+n+mn}\,
\frac{e^{- \frac{\pi}{2\tau_2} \left|\la\right|^2}}{\la^2} 
\equiv - \sum_{\la \in \La'} \, (-1)^{m+n}\, \frac{q^{\frac{1}{2} m^2} e^{- \frac{\pi}{2\tau_2} \la^2} }
{\la^2}  .
\label{def g}
\end{equation}
Then, we obtain 
\begin{eqnarray}
&& \frac{\del}{\del \bar{\tau}} g(\tau) = - \sum_{\la \in \La'} \, (-1)^{m+n+mn}\,
e^{- \frac{\pi}{2\tau_2} \left|\la\right|^2} \, \frac{i \pi }{4\tau_2^2}
=  \frac{i \pi}{4 \tau_2^2} = - \frac{1}{2} \frac{\del}{\del \bar{\tau}} \left(\frac{\pi}{\tau_2}\right).
\label{eval g 1}
\end{eqnarray}
Here we again used \eqn{identity 1}.
By the definition \eqn{def g}, 
$g(\tau)$ should be a non-holomorphic modular form of weight 2 for $\Gamma(1)$. 
Thus  $g(\tau)$ is uniquely determined by \eqn{eval g 1} as
\begin{equation}
g(\tau) = \frac{1}{2} \hG_2(\tau) .
\label{eval g 2}
\end{equation} 


Combining \eqn{def g}, \eqn{eval g 2} with \eqn{eval hH2 0}, we finally obtain
\begin{eqnarray}
\hH^{(2)}(\tau) & = &
\frac{1}{4\pi^2}
\left[
 \sum_{\la \in \La'} \, (-1)^{m+n+mn}\,\frac{ 2\pi i m e^{- \frac{\pi}{2\tau_2} \left|\la\right|^2 }}{\la}  
+ \hG_2(\tau) + \frac{\pi}{\tau_2} 
\right]
\nn
& = & \frac{1}{4\pi^2}
\left[
 \sum_{\la \in \La'} \, (-1)^{m+n+mn}\,\frac{ 2\pi i m e^{- \frac{\pi}{2\tau_2} \left|\la\right|^2 }}{\la}  
+ G_2(\tau) 
\right],
\label{result hH2}
\end{eqnarray} 
which is the second line of  \eqn{formula hH2}.

~


\section*{Appendix C: ~ A Lemma for Weak Jacobi Form }
\label{lemma}

\setcounter{equation}{0}
\def\theequation{C.\arabic{equation}}

In this appendix, we present a simple lemma that is necessary for the proof of 
\eqn{main id hf hF}.

~


\noindent
{\bf Lemma:}

Let $\Phi(\tau,z)$ be any holomorphic weak Jacobi form of weight 0 and index $d (\in \bz_{>0})$.
Then,  the `NS-counterpart'
$
\Phi^{(\sNS)}(\tau,z) \equiv q^{\frac{d}{4}} y^d \, \Phi\left(\tau, z + \frac{\tau+1}{2} \right) 
$
has the following behavior around $\tau \sim i\infty$ (for a generic value of $z$);
\begin{equation}
\Phi^{(\sNS)}(\tau,z)\sim q^{\al}, 
\hspace{7mm} - \frac{d}{4} \leq \al \leq \frac{d}{12},
\hspace{1cm} (\tau \sim i\infty).
\end{equation}

~


\noindent
{\bf [proof]}

The proof is just straightforward. Let us recall that bases of 
holomorphic weak Jacobi forms of weight 0 and index $d (\in \bz_{>0})$
consist of the functions defined by 
\begin{eqnarray}
&& \phi_{(n_1, n_2, n_3)} (\tau,z) \propto \sum_{\sigma \in S_3}\,
\left(\frac{\th_2(\tau,z)}{\th_2(\tau)}\right)^{2n_{\sigma(1)}}
\left(\frac{\th_3(\tau,z)}{\th_3(\tau)}\right)^{2n_{\sigma(2)}}
\left(\frac{\th_4(\tau,z)}{\th_4(\tau)}\right)^{2n_{\sigma(3)}},
\nn
&& \phi_{(n_1, n_2, n_3)} (\tau,0) =1, 
\hspace{1cm} 
\any (n_1, n_2, n_3) \in \cS^{(d)}, 
\end{eqnarray}
with 
\begin{equation}
\cS^{(d)} :=  
\left\{ (n_1, n_2, n_3) \in \bz_{\geq 0}^3 ~; ~ n_1 \geq n_2 \geq n_3 \geq 0, ~
n_1 + n_2 + n_3 = d \right\}. 
\end{equation}


Now, 
we define the NS counterpart of the function
$\phi_{(n_1, n_2, n_3)} (\tau,z)$ by the spectral flow $z\, \mapsto \, z+ \frac{\tau+1}{2}$;
$$
 \phi^{(\sNS)}_{(n_1,n_2,n_3)} (\tau,z) : = q^{\frac{d}{4}} y^{d}\, 
 \phi_{(n_1,n_2,n_3)} \left(\tau,z+ \frac{\tau+1}{2}\right).
$$
Then, by using the evaluations around $\tau \sim i\infty$
$$
\left(\frac{\th_4(\tau,z)}{\th_2(\tau)}\right)^{2} \sim q^{-\frac{1}{4}}, \hspace{1cm}
\left(\frac{\th_1(\tau,z)}{\th_3(\tau)}\right)^{2} \sim q^{\frac{1}{4}}, \hspace{1cm}
\left(\frac{\th_2(\tau,z)}{\th_4(\tau)}\right)^{2} \sim q^{\frac{1}{4}}, 
$$
we can easily find that the behavior of $\phi^{(\sNS)}_{(n_1,n_2,n_3)} (\tau,z)$ becomes 
\begin{equation}
\phi^{(\sNS)}_{(n_1,n_2,n_3)} (\tau,z) \sim q^{-\frac{1}{4}n_1 + \frac{1}{4} (n_2+n_3)} 
= q^{-\frac{1}{2} \left(n_1 - \frac{d}{2}\right)}, 
\hspace{1cm} (\tau \, \sim \, i\infty).
\end{equation}
Since we are assuming $n_1+n_2+n_3 =d$ and $n_1 \geq n_2 \geq n_3 \geq 0$, 
we find 
$$
 \frac{d}{3} \leq n_1 \leq d.
$$
Therefore, we obtain
\begin{eqnarray}
&& \phi^{(\sNS)}_{(n_1,n_2,n_3)} (\tau,z) \sim q^{\al}, \hspace{1cm}
- \frac{d}{4} \leq \al \leq \frac{d}{12}, \hspace{1cm}
 (\tau \sim i\infty),
 \label{IR behavior phi NS}
\end{eqnarray}
for  $\any (n_1,n_2,n_3) \in \cS^{(d)}$.
\hspace{5mm} 
{\bf (Q.E.D)}

~



\section*{Appendix D : ~ More Comments on Eq. \eqn{q-exp HN}}

\setcounter{equation}{0}
\def\theequation{D.\arabic{equation}}


In this appendix\footnote
{Appendix D is based on the paper: 'Addendum to '$\cN=4$ Liouville Theory and Moonshine Phenomena'\,', which will be published in PTEP.}
 we exhibit some computations with respect to the function $H^{(N)}(\tau)$ 
defined in \eqn{q-exp HN},
which is a mock modular form of weight 2 and  plays an important role in the duality of Mathieu and  umbral moonshines \cite{EOT,umbral1,umbral2}. 
This has been derived  based on the elliptic genus of the  $\cN=4$ superconformal system with $\hat{c} \left( \equiv \frac{c}{3} \right) =2 $; 
$$
\left[ SU(2)_N/U(1) \otimes SL(2)_N/U(1) \right]_{\bz_N\msc{-orbifold}}, 
$$
computed in  \cite{ES-BH,CH}, 
and written explicitly  as follows
($q\equiv e^{2\pi i \tau}$, $N \in \bz_{> 0}$) ; 
\begin{eqnarray}
H^{(N)}(\tau) & =& 
\frac{N-1}{12} - 2N \sum_{n=1}^{\infty} \, \frac{n q^n}{1-q^n} 
+  2 \sum_{m=1}^{\infty} \, q^{Nm^2}  \left[ \frac{q^{m}}{\left(1 -q^m \right)^2}
+ Nm \frac{1+q^m}{1-q^m} \right],
\label{q-exp HN app}
\end{eqnarray}
For instance,  we obtain for $N=2,3,4$,
\begin{eqnarray}
&& -H^{(2)}(\tau) 
= -\frac{1}{12} + 4 q + 8q^2+6q^3+16q^4+10q^5+ 32q^6 + O(q^7),
\nn
&& -H^{(3)}(\tau)
= -\frac{1}{6} + 6q+18q^2+18q^3+28q^4+20q^5+54 q^6+ O(q^7),
\nn
&& -H^{(4)}(\tau)
=  -\frac{1}{4} + 8q+24q^2+32q^3+48q^4+30q^5+76q^6 + O(q^7),
\label{ex HN app}
\end{eqnarray}


On the other hand, the formula \eqn{rel hhX} is written as 
\begin{eqnarray}
h^X(\tau) &=& \sum_{r=1}^{N-1} \, h_r^X(\tau) \chi^{(N-2)}_{r-1}(\tau,0)
\equiv 
 \frac{1}{\eta(\tau)^3} \, \sum_{r=1}^{N-1}\, h^X_r(\tau) S_{r,N}(\tau).
\label{h rel app}
\end{eqnarray}
Here, '$X$' expresses a simply-laced root system of rank 24   which possesses  the common Coxetor number $N$ and labels  each Niemeier lattice.   
$\chi^{(N-2)}_{\ell}(\tau,0)$ is the character of affine $\widehat{SU}(2)_{N-2}$ with isospin $\ell/2$ and  
$S_{r,N}(\tau)$ denotes the `unary theta function' defined by
\begin{equation}
S_{r,N}(\tau) := \left. \frac{1}{2\pi i} \frac{\del}{\del z} \Th{r}{N}(\tau,2z) \right|_{z=0}
\equiv \sum_{n\in \bz}\,(r+2Nn) \, q^{\frac{1}{4N} (r+2Nn)^2}.
\label{def SrN app}
\end{equation} 
The explicit $q$-expansions of $S_{r,N}(\tau)$ for $N=2,3,4$ are given as 
\begin{eqnarray}
&& S_{1,2}(\tau) = \eta(\tau)^3,
\nn
&& S_{1,3}(\tau) = q^{\frac{1}{12}} \left[ 1- 5 q^2 + 7 q^4 - 11 q^{10} + \cdots \right]
\nn
&& S_{2,3}(\tau) = 2 q^{\frac{1}{3}} \left[1-2 q + 4 q^5 - 5 q^8 + \cdots \right]
\nn
&& S_{1,4}(\tau) = q^{\frac{1}{16}} 
\left[1- 7 q^3+ 9 q^5 +\cdots  \right],
\nn
&& S_{2,4} (\tau) = 2 q^{\frac{1}{4}} \left[1 - 3 q^2 + 5 q^6 +\cdots \right],
\nn
&& S_{3,4} (\tau) = q^{\frac{9}{16}}
\left[ 3 - 5 q + 11 q^7 + \cdots \right].
\label{ex SrN app}
\end{eqnarray}
In the cases when $X$ only includes $A$-type root system, that is, 
$$
X = \left(A_{N-1}\right)^M, \hspace{1cm} M \equiv \frac{24}{N-1}, ~~ (\mbox{$N-1$ divides 24.})
$$
we shall use the abbreviated notations;
$
h^{(N)}(\tau) \equiv h^X(\tau) ,
$ 
$
h_r^{(N)}(\tau) \equiv h_r^X(\tau) .
$
In this case, we can express $h^{(N)}(\tau)$ in terms of $H^{(N)}(\tau)$ as follows;
\begin{equation}
h^{(N)} = - \frac{24}{N-1} \frac{H^{(N)}(\tau)}{\eta(\tau)^3}.
\label{hN app}
\end{equation}
Thus, the formula \eqn{h rel app} is equivalent with 
\begin{equation}
- \frac{24}{N-1} H^{(N)}(\tau) = \sum_{r=1}^{N-1}\, h^{(N)}_r(\tau) S_{r,N}(\tau).
\label{h rel 2 app}
\end{equation}
The $q$-expansions of 
$h^{(N)}_r$ for $N=2,3,4$ are presented in \cite{umbral1} (page 28, eqs. (2.65)-(2.67)); 
\begin{eqnarray}
&& h^{(2)}_1(\tau) = 2 q^{-\frac{1}{8}} \left[-1 + 45 q + 231 q^2 + 770 q^3 + 2277 q^4 + 5796 q^5 
\cdots \right] ,
\nn
&& h^{(3)}_1(\tau) = 2 q^{-\frac{1}{12}} \left[-1 + 16 q + 55 q^2 + 144 q^3
+ 330q^4 + 704 q^5 + \cdots \right] ,
\nn
&& h^{(3)}_2(\tau) = 2 q^{\frac{2}{3}} \left[
10+44q+110q^2+280q^3+ 572q^4 + 1200q^5 + \cdots
\right],
\nn
&& h^{(4)}_1(\tau) = 2 q^{-\frac{1}{16}} 
\left[
-1+7q+21 q^2 + 43 q^3 + 94 q^4 + 168 q^5 + \cdots
\right],
\nn
&& h^{(4)}_2(\tau) = 2 q^{\frac{3}{4}}
\left[
8+24 q + 56 q^2 + 112 q^3 + 216 q^4 + 392 q^5 + \cdots
\right],
\nn
&& h^{(4)}_3(\tau) = 2 q^{\frac{7}{16}}
\left[3 + 14 q + 28 q^2 + 69q^3 + 119 q^4 + 239 q^5 + \cdots\right].
\label{hNr ex app}
\end{eqnarray}

We can numerically check the identity \eqn{h rel app} or \eqn{h rel 2 app} for $N=2,3,4$ 
by using \eqn{ex HN app}, \eqn{ex SrN app}, and \eqn{hNr ex app};
\begin{itemize}
\item{\bf $X= A^{24}_1$ :}

\eqn{h rel app} reduces to a trivial identity 
$h^{(2)}(\tau)= h^{(2)}_1(\tau) $ because $S_{1,2}(\tau) = \eta(\tau)^3$ holds,
and \eqn{hN app} gives the well-known formula  of Mathieu moonshine \cite{EOT};
\begin{eqnarray}
h^{(2)}(\tau) &=& -24 \frac{H^{(2)}(\tau) }{\eta(\tau)^3} 
\nn
&=& 
2 q^{-\frac{1}{8}}\, \left[-1+ 45 q + 231 q^2 + 770 q^3 + 2277 q^4 + 5796 q^5 
+ \cdots \right].
\label{Mathieu}
\end{eqnarray}
The second line has been numerically checked.

\item $X=A_2^{12}$ :

\eqn{h rel 2 app} reduces to 
\begin{eqnarray}
-12 H^{(3)}(\tau) & =& h^{(3)}_1(\tau) S_{1,3}(\tau) + h^{(3)}_2(\tau) S_{2,3}(\tau)
\nn
& =& -2 + 72 q + 216 q^2 +216 q^3 + 336 q^4 + \cdots ,
\label{check A2}
\end{eqnarray}
and the second line is numerically checked. 

\item $X=A_3^8$ :

\eqn{h rel 2 app} similarly reduces to the next identity that is numerically checked;
\begin{eqnarray}
-8 H^{(4)}(\tau) & =& h^{(4)}_1(\tau) S_{1,4}(\tau) + h^{(4)}_2(\tau) S_{2,4}(\tau)
+ h^{(4)}_3(\tau) S_{3,4}(\tau)
\nn
& =& -2 + 64 q + 192 q^2 +256 q^3 + 384 q^4 + \cdots .
\label{check A3}
\end{eqnarray}

\end{itemize}

~


We  also note that the function $H^{(N)}(\tau)$  
should be related to the `second helicity supertrace' (supersymmetric index)
computed in \cite{HM,HMN} based on a different string-theoretical construction. 
Indeed, one finds the `number theoretical' formula for this index; 
\begin{align}
& \chi_2^{(k,d)}(\tau) = \left(\frac{k}{d} -d \right) E_2(\tau) - 24 \cF^{(k,d)}_2(\tau),
\label{formula chi2}
\\
&\cF^{(k,d)}_2(\tau) = \left(d \sum_{\stackrel{r,s}{kr> d^2 s >0}} -\frac{k}{d}  \sum_{\stackrel{r,s}{d^2 r> k s >0}}  \right) s q^{rs},
\label{formula Fkd}
\end{align} 
on page 12 of \cite{HMN} (derived in ref.[57] of \cite{HMN}, more precisely),
where $E_2(\tau)$ denotes the normalized 2nd Eisenstein series;
$$
E_2(\tau) \left(\equiv \frac{3}{\pi^2} G_2(\tau) \right) = 1 - 24 \sum_{n=1}^{\infty}\, \frac{n q^n}{1-q^n}.
$$
The precise relation between $H^{(N)}(\tau)$ and the index $\chi_2$ is written as 
\begin{equation}
H^{(N)}(\tau) = \frac{1}{12} \chi_2^{(N,1)}(\tau).
\label{id HN chi2} 
\end{equation} 
One can confirm this identity by focusing on the mock modularity and the coincidence of shadow. 
We here present an elementary proof by directly evaluating the both sides of \eqn{id HN chi2}.


In fact, 
\eqn{q-exp HN app} can be rewritten as
 ($y\equiv e^{2\pi i w}$) 
\begin{align}
H^{(N)}(\tau) = \frac{N}{12} E_2(\tau) - \frac{1}{12} + \frac{1}{2\pi i} \left. \frac{\del}{\del w} \sum_{m\neq 0} \, \left[ \frac{y^{2N m} q^{N m^2}}{1-y q^m}
- \frac{1}{2} y^{2N m} q^{N m^2}
\right]\right|_{w=0},
\label{proof HN chi2 1}
\end{align}
and since we have
\begin{align}
\sum_{m\neq 0} \, \frac{y^{2N m} q^{N m^2}}{1-y q^m}
& = \sum_{m=1}^{\infty} \, \left[ \frac{y^{2N m} q^{N m^2}}{1-y q^m} -  \frac{y^{-2N m} q^{N m^2}}{1-y^{-1} q^m} + y^{-2Nm}q^{Nm^2} \right]
\nn
& = \sum_{m=1}^{\infty} \, \sum_{n=0}^{\infty} \, \left(y^{2Nm+n} - y^{-(2Nm+n)}\right) q^{Nm^2 + mn} + \sum_{m=1}^{\infty}\, y^{-2Nm} q^{Nm^2},
\end{align}
we obtain 
\begin{align}
H^{(N)}(\tau) & = \frac{N}{12} E_2(\tau) - \frac{1}{12}  
+ 2 \sum_{m=1}^{\infty}\, \sum_{n=0}^{\infty}\, (2Nm+n) q^{Nm^2 + mn} - \sum_{m=1}^{\infty} \, 2Nm q^{Nm^2}
\nn
&  = \frac{N-1}{12} - 2N \sum_{n=1}^{\infty}\, \frac{n q^n}{1-q^n}  
+ 2 \sum_{m=1}^{\infty}\, \sum_{n=1}^{\infty}\, (2Nm+n) q^{m(Nm+n)} + \sum_{m=1}^{\infty} \, 2Nm q^{Nm^2}.
\label{proof HN chi2 2}
\end{align}
On the other hand, $\cF^{(N,1)}_2(\tau)$ given in \eqn{formula Fkd} is rewritten as 
\begin{align}
\cF^{(N,1)}_2(\tau) & = \left(\sum_{\stackrel{r,s}{Nr>  s >0}} - N  \sum_{\stackrel{r,s}{r>  N s >0}}  \right) s q^{rs}
\nn
& = \sum_{r=1}^{\infty}\, \sum_{s=1}^{Nr-1}\, s q^{rs}
- N \sum_{s=1}^{\infty}\, \sum_{a=1}^{\infty}\, s q^{s(Ns+a)}
\nn
& = \sum_{r=1}^{\infty}\, \sum_{s=1}^{\infty}\, s q^{rs} - \sum_{r=1}^{\infty}\, \sum_{a=0}^{\infty}\, (Nr+a) q^{r(Nr+a)}
- N \sum_{s=1}^{\infty}\, \sum_{a=1}^{\infty}\, s q^{s(Ns+a)}
\nn
& = \sum_{n=1}^{\infty} \, \frac{n q^n}{1-q^n} - \sum_{m=1}^{\infty}\, \sum_{n=1}^{\infty}\, (2 N m+n) q^{m(Nm+n)} - \sum_{m=1}^{\infty}\, Nm q^{Nm^2}.
\end{align}
Thus we obtain
\begin{align}
\frac{1}{12} \chi_2^{(N,1)} (\tau) & \equiv \frac{N-1}{12} E_2(\tau) - 2 \cF^{(N,1)}_2 (\tau) 
\nn
 & = \frac{N-1}{12} - 2(N-1) \sum_{n=1}^{\infty} \, \frac{n q^n}{1-q^n}
 \nn
 & 
 \hspace{1cm}
-2  \sum_{n=1}^{\infty} \, \frac{n q^n}{1-q^n} +2  \sum_{m=1}^{\infty}\, \sum_{n=1}^{\infty}\, (2 N m+n) q^{m(Nm+n)} + 2 \sum_{m=1}^{\infty}\, Nm q^{Nm^2}
\nn
& = \frac{N-1}{12} - 2N \sum_{n=1}^{\infty} \, \frac{n q^n}{1-q^n}
+2  \sum_{m=1}^{\infty}\, \sum_{n=1}^{\infty}\, (2 N m+n) q^{m(Nm+n)} + 2 \sum_{m=1}^{\infty}\, Nm q^{Nm^2},
\label{proof HN chi2 3}
\end{align}
which coincides with \eqn{proof HN chi2 2} as expected.


\newpage


\end{document}